\documentclass[usenatbib,useAMS,usedcolumn]{mn2e}
\usepackage{times}
\usepackage{psfig}
\usepackage{epsfig}
\usepackage{graphicx}
\def\arcm{\hbox{$^\prime$}}
\def\eg{{\it e.g.\ }}
\def\etc{{\it etc.\ }}
\def\ie{{\it i.e.\ }}
\def\cf{{\it c.f.\ }}
\def\deg{\hbox{$^\circ$}}
\def\spose#1{\hbox to 0pt{#1\hss}}
\def\simpropto{$\mathrel{\spose{\lower 3pt\hbox{$\sim$}}
        \raise 2.0pt\hbox{$\propto$}}$\thinspace}
\newcommand{\rosat}{\emph{ROSAT}}
\newcommand{\chandra}{\emph{Chandra}}
\newcommand{\xmm}{\emph{XMM-Newton}}
\newcommand{\asca}{\emph{ASCA}}
\newcommand{\einstein}{\emph{Einstein}}
\newcommand{\Lx}{\ensuremath{L_{\mathrm{X}}}}
\newcommand{\Tx}{\ensuremath{T_{\mathrm{X}}}}
\newcommand{\Zsol}{\ensuremath{Z_{\odot}}}
\newcommand{\Lsol}{\ensuremath{L_{\odot}}}
\newcommand{\Msol}{\ensuremath{M_{\odot}}}
\newcommand{\LB}{\ensuremath{L_{B}}}

\newcommand{\LT}{\ensuremath{\mbox{\Lx :\Tx}}}
\newcommand{\LxLb}{\ensuremath{\mbox{\Lx/\LB}}}
\newcommand{\LxLbtwo}{\ensuremath{\mbox{\Lx :\LB}}}
\newcommand{\LbT}{\ensuremath{\mbox{\LB :\Tx}}}
\newcommand{\Bfit}{\ensuremath{\beta_{fit}}}
\newcommand{\BfitT}{\ensuremath{\mbox{\Bfit :\Tx}}}
\newcommand{\sigT}{\ensuremath{\mbox{$\sigma$:\Tx}}}
\newcommand{\ST}{\ensuremath{\mbox{S :\Tx}}}
\newcommand{\NH}{\ensuremath{n_{\mathrm{H}}}}
\newcommand{\Fgas}{\ensuremath{\mathit{f}_{gas}}}
\newcommand{\s}{\ensuremath{\mbox{~s}}}
\newcommand{\ps}{\ensuremath{\s^{-1}}}
\newcommand{\km}{\ensuremath{\mbox{~km}}}
\newcommand{\Mpc}{\ensuremath{\mbox{~Mpc}}}
\newcommand{\pMpc}{\ensuremath{\Mpc^{-1}}}
\newcommand{\kmpspMpc}{\ensuremath{\km \ps \pMpc\,}}
\newcommand{\erg}{\ensuremath{\mbox{~erg}}}
\newcommand{\ergps}{\ensuremath{\erg \ps}}
\newcommand{\kmps}{\ensuremath{\km \ps}}

\begin{document}

\title[
X--ray Scaling Properties of Early--type Galaxies 
]
{
X--ray Scaling Properties of Early--type Galaxies
}
\author[
Ewan O'Sullivan et al. 
]
{
Ewan O'Sullivan\footnotemark, Trevor J. Ponman, Ross S. Collins\\
School of Physics and Astronomy, 
University of Birmingham, Edgbaston, Birmingham B15 2TT \\
\\
}

\date{Accepted 2003 ??. Received 2002 ??; in original form 2002 ??}
\pagerange{\pageref{firstpage}--\pageref{lastpage}}
\def\LaTeX{L\kern-.36em\raise.3ex\hbox{a}\kern-.15em
    T\kern-.1667em\lower.7ex\hbox{E}\kern-.125emX}

\label{firstpage}

\maketitle

\begin{abstract}
  We present an analysis of 39 X--ray luminous early--type galaxies
  observed with the \rosat\ PSPC. Using multi--component spectral and
  spatial fits to these data we have measured halo abundance, temperature,
  luminosity and surface brightness profile. We compare these measurements
  to similar results from galaxy groups and clusters, fitting a number of
  relations commonly used in the study of these larger objects. In
  particular, we find that the \sigT\ relation for our sample is similar to
  that reported for clusters, consistent with $\beta_{spec}=1$, and that
  the \LT\ relation has a steep slope (gradient 4.8$\pm$0.7) comparable
  with that found for galaxy groups.  Assuming isothermality, we construct
  3-dimensional models of our galaxies, allowing us to measure gas entropy.
  We find no correlation between gas entropy and system mass, but do find a
  trend for low temperature systems to have reduced gas fractions. We
  conclude that the galaxies in our sample are likely to have developed
  their halos through galaxy winds, influenced by their surrounding
  environment.
\end{abstract}


\footnotetext{\textit{Present address}: Harvard-Smithsonian
  Center for Astrophysics, 60 Garden Street, Cambridge, MA 02138, USA\\
Email: ejos@head-cfa.harvard.edu}

\section{Introduction}
Early-type galaxies have been known to possess large halos of hot gas since
the detection by \einstein\ of X--ray emission from the elliptical
population in the Virgo cluster \citep{Formanetal79}. Successive
generations of X--ray observatories have been used to observe these
galaxies, and the advent of \xmm\ and \chandra\ has allowed the complex
nature of their emission to be studied in detail. Most of the work in this
area has focused on the various sources of emission within early--type
galaxies (hot gas, X--ray binaries, AGN) and on the surprisingly
complicated relation between optical and X--ray luminosity. However, at a
quite fundamental level, early--type galaxies resemble the groups and
clusters in which they typically reside. Simulations of dark matter halos
suggest that they have similar profiles at all mass scales
\citep{NavarroFW97}. If we consider clusters, groups and galaxies as
potentials containing hot gas, we might expect the properties of the halos
to be similar across a wide range of masses.

A comparison between halos on different scales becomes increasingly
interesting considering the importance of entropy changes in governing the
behaviour of group and cluster halos. Observations of galaxy groups have
demonstrated that these systems do not behave as might be expected from
scaling clusters, but instead require non-gravitational processes. One of
the clearest signs of this is the entropy floor \citep{Ponmanetal99}.
Whereas in more massive systems gas entropy scales with the total mass, in
groups it appears to reach a roughly constant minimum level. A number of models
have been put forward to explain this behaviour, including raising the
entropy through the injection of energy by AGN \citep{Wuetal00} or star
formation \citep{Ponmanetal99}, or through the radiative cooling and removal
of low entropy gas \citep{Muaonwongetal01}. It is notable that for any of
these processes, the origin of the entropy rise would 
be related to the galaxies in the system. As early--type galaxies
possess their own halos, we might expect to see evidence of these processes
in their X--ray properties, and for the effect to be strongest in these
systems, owing to their position at the bottom of the mass scale.

There is also evidence from previous studies of early--type galaxies
\citep{Helsdonetal01,OFP01cat} that galaxies in the centres of X--ray
bright groups are affected by their environment. They have a significantly
steeper \LxLbtwo\ relation, and are on average considerably more luminous
than normal ellipticals. A large fraction of group dominant galaxies in one
sample have been shown to have temperature profiles indicative of central
cooling \citep{Helsdonetal01}, leading to the suggestion that their halos
are actually the product of cooling flows associated with the surrounding
group. Considering the differences between these dominant galaxies and
their more normal counterparts, and the biasing effect their inclusion in
samples of early--type galaxies seems to have, further investigation of the
processes which have shaped their halos seems warranted.

We have compiled a sample of 39 large, X--ray luminous early--type
galaxies for which there is good quality \rosat\ PSPC data available. We
have analysed these data, and fitted two dimensional, two component surface 
brightness profiles to them. We have also fitted two component spectral
models, temperature, abundance and hardness profiles and produced three
dimensional models of the galaxies. These allow us to model out
contamination from surrounding cluster or group emission and the discrete
source population within the galaxy. We can therefore examine the
properties of the halo in detail, for the first time in a sample of this
size. We can also compare the behaviour of this sample to that of samples
of groups and clusters through relations between parameters such as
temperature, optical and X--ray luminosity, velocity dispersion, surface
brightness slope and gas entropy. In most cases, this is the first time
these relations have been studied for halos at this mass scale.

The paper is organised as follows. In Section~\ref{sample} we describe our
sample and the selection criteria used to create it. Section~\ref{datared}
gives details of the techniques used in reduction of the \rosat\ PSPC data,
and Section~\ref{analysis} describes the spectral and spatial fitting
processes. Our results are presented in Section~\ref{sec:results}, with
data from clusters and groups of galaxies included for comparison. We
discuss the results and their implications in Section~\ref{sec:discuss},
and give our conclusions in Section~\ref{sec:conc}. Throughout the paper we
assume H$_0$=50 \kmpspMpc, in order to simplify comparison with previous
studies of groups and clusters. Optical luminosities are normalised using
the solar luminosity in the $B$ band, $L_{B\odot}$ = 5.2$\times$10$^{32}$
\ergps.

\section{Sample Selection}
\label{sample}
Our sample was selected from the Lyon-Meudon Extragalactic Data Archive
(LEDA), specifically the PGC-ROM 1996 (2$^{\rm nd}$ edition). This contains
information on $\sim$100,000 galaxies, of which $\sim$40,000 have the
necessary redshift and morphological data. Galaxies were selected to match
the following selection criteria:
\begin{enumerate}
\renewcommand{\theenumi}{(\arabic{enumi})}
\item Absolute magnitude M$_B <$ --19
\item Morphological T-type $<$ --2
\item Virgocentric flow corrected recession velocity V$_{rec} <$ 10,500 km
  s$^{-1}$\end{enumerate}
These criteria were chosen to produce a selection of optically luminous
nearby early--type galaxies.  

A list of galaxies matching these criteria was then compared to a catalogue
of {\it ROSAT} PSPC pointings, to produce an initial sample of galaxies
with X--ray data. Only galaxies lying within the PSPC support structure
(\ie within $\sim$30\arcm\ of the pointing) were accepted, so as to ensure
that the X--ray data were not strongly affected by vignetting effects or
off--axis resolution problems. Pointings of less than 10 ksec were also
ignored, as these were unlikely to provide X--ray data of sufficient
quality. This initial sample contained 47 galaxies.

We then examined images of the raw X--ray data for each galaxy, to look for
potential problems. In some cases we found that the galaxies appeared to be
extremely compact or point--like, suggesting that surface brightness
fitting would be difficult or impossible. These objects were removed from
the sample, as were galaxies in which an AGN or nearby quasar dominated the
X-ray emission, to produce a final sample of 39 X--ray luminous early-type
galaxies. It is worth noting that as the fraction of galaxies in which the
halo was too compact or faint for analysis was small ($\sim$4\%), 
it appears that the
majority of massive early--type galaxies do possess bright, extended X--ray
halos.  Table~\ref{tab:basic} lists our targets.

\begin{table*}
\begin{center}
\begin{tabular}{lcccccccc}
\hline
Name & RA & DEC &$\sigma$ & V$_{rec}$ & D & R$_e$ & T & Environment \\
 & (2000) & (2000) & (km s$^{-1}$) & (km s$^{-1}$) & (Mpc) & ($\arcmin$) & & \\
\hline 
ESO 443-24 & 13 01 01.6 & -32 26 20 & 279.9 & 4970.1 & 99.4 & 0.388 & -3.1 & BGG \\ 
IC 1459 & 22 57 09.5 & -36 27 37 & 321.4 & 1522.0 & 28.3 & 0.644 & -4.7 & BGG \\ 
IC 4296 & 13 36 38.8 & -33 57 59 & 341.2 & 3587.9 & 71.8 & 0.953 & -4.8 & BGG \\ 
IC 4765 & 18 47 19.0 & -63 19 49 & 288.4 & 4344.9 & 86.9 & 0.239 & -3.9 & BGG \\ 
NGC 499 & 01 23 11.5 & +33 27 36 & 264.2 & 4482.7 & 82.8 & 0.346 & -2.9 & BGG \\ 
NGC 507 & 01 23 40.0 & +33 15 22 & 295.8 & 5015.7 & 100.3 & 1.285 & -3.3 & BGG \\ 
NGC 533 & 01 25 31.4 & +01 45 35 & 250.0 & 5411.4 & 95.5 & 0.792 & -4.7 & BGG \\ 
NGC 720 & 01 53 00.4 & -13 44 21 & 237.7 & 1622.0 & 31.2 & 0.659 & -4.7 & BGG \\ 
NGC 741 & 01 56 20.9 & +05 37 44 & 288.4 & 5529.9 & 91.6 & 0.869 & -4.8 & BGG \\ 
NGC 1332 & 03 26 17.3 & -21 20 09 & 328.1 & 1355.8 & 29.5 & 0.467 & -2.9 & Group \\ 
NGC 1380 & 03 36 26.9 & -34 58 33 & 240.4 & 1617.6 & 27.2 & 0.659 & -2.3 & Cluster \\ 
NGC 1395 & 03 38 29.6 & -23 01 40 & 241.0 & 1516.1 & 30.8 & 0.757 & -4.8 & BGG \\ 
NGC 1399 & 03 38 28.9 & -35 26 58 & 329.6 & 1211.2 & 27.2 & 0.706 & -4.5 & BCG \\ 
NGC 1404 & 03 38 51.7 & -35 35 36 & 212.3 & 1701.9 & 27.2 & 0.446 & -4.7 & Cluster \\ 
NGC 1407 & 03 40 12.3 & -18 34 52 & 279.3 & 1612.0 & 30.9 & 1.199 & -4.5 & BGG \\ 
NGC 1549 & 04 15 45.0 & -55 35 31 & 203.2 &  932.5 & 21.7 & 0.792 & -4.3 & Group \\ 
NGC 1553 & 04 16 10.3 & -55 46 51 & 167.5 &  805.8 & 21.7 & 1.094 & -2.3 & BGG \\ 
NGC 2300 & 07 32 19.6 & +85 42 32 & 263.0 & 2249.7 & 41.5 & 0.524 & -3.4 & BGG$^{\dag}$ \\ 
NGC 2832 & 09 19 46.5 & +33 45 02 & 341.2 & 6992.2 & 128.9 & 0.426 & -4.3 & BGG \\ 
NGC 3091 & 10 00 13.8 & -19 38 14 & 303.4 & 3670.2 & 76.2 & 0.512 & -4.7 & BGG \\ 
NGC 3607 & 11 16 54.1 & +18 03 12 & 216.8 &  999.6 & 29.7 & 1.094 & -3.1 & BGG \\ 
NGC 3923 & 11 51 02.1 & -28 48 23 & 269.8 & 1468.0 & 26.8 & 0.889 & -4.6 & BGG \\ 
NGC 4073 & 12 04 26.5 & +01 53 48 & 267.9 & 5970.6 & 119.1 & 0.931 & -4.1 & BGG \\ 
NGC 4125 & 12 08 07.1 & +65 10 22 & 239.9 & 1618.6 & 38.9 & 0.998 & -4.8 & BGG \\ 
NGC 4261 & 12 19 22.7 & +05 49 36 & 316.2 & 2244.0 & 47.2 & 0.644 & -4.8 & Cluster \\ 
NGC 4291 & 12 20 18.1 & +75 22 21 & 287.7 & 2043.9 & 36.8 & 0.245 & -4.8 & Group \\ 
NGC 4365 & 12 24 27.9 & +07 19 06 & 268.5 & 1290.4 & 23.9 & 0.830 & -4.8 & Cluster \\ 
NGC 4472 & 12 29 46.5 & +07 59 58 & 304.8 &  931.8 & 23.9 & 1.734 & -4.7 & BCG \\ 
NGC 4552 & 12 35 39.9 & +12 33 25 & 264.2 &  372.3 & 23.9 & 0.500 & -4.6 & Cluster/AGN \\ 
NGC 4636 & 12 42 49.8 & +02 41 17 & 211.3 & 1125.2 & 23.9 & 1.694 & -4.8 & Cluster/BGG \\ 
NGC 4649 & 12 43 40.2 & +11 32 58 & 342.8 & 1221.9 & 23.9 & 1.227 & -4.6 & Cluster \\ 
NGC 4697 & 12 48 35.9 & -05 48 02 & 173.4 & 1232.2 & 22.7 & 1.256 & -4.7 & BGG \\ 
NGC 5128 & 13 25 29.0 & -43 01 00 & 142.6 &  385.6 &  5.8 & 0.708 & -2.1 & BGG/AGN \\ 
NGC 5322 & 13 49 15.5 & +60 11 29 & 239.9 & 2035.5 & 41.7 & 0.587 & -4.8 & BGG \\ 
NGC 5419 & 14 03 38.6 & -33 58 41 & 329.6 & 4027.5 & 80.5 & 0.723 & -4.2 & BGG$^{\dag}$ \\ 
NGC 5846 & 15 06 29.3 & +01 36 25 & 250.0 & 1890.0 & 34.4 & 1.377 & -4.7 & BGG \\ 
NGC 6269 & 16 57 58.4 & +27 51 19 & 224.4 & 10435.0 & 208.7 & 0.574 & -4.8 & BGG \\ 
NGC 6482 & 17 51 49.0 & +23 04 20 & 302.0 & 4102.0 & 82.0 & 0.132 & -4.8 & Field ? \\ 
NGC 7619 & 23 20 14.7 & +08 12 23 & 310.5 & 3825.5 & 60.0 & 0.536 & -4.7 & BCG \\ 
\hline
\end{tabular}
\end{center}
\caption{\label{tab:basic}
A list of galaxies included in our sample. RA and DEC are taken from the
LEDA catalogue, as are morphological type (T), Recession velocity (which is 
corrected for Virgocentric flow and movement within the local group), and
velocity dispersion, $\sigma$. Distances are taken from
\protect\citet{PrugnielSimien96} where possible, or calculated from the
recession velocity. H$_0$=50 is assumed in both cases, and for most
objects, the distance used is that of the group or cluster in which is resides. Optical
effective radii, R$_e$, are taken from \protect\citet{Faberetal89} and
\protect\citet{PrugnielHeraudeau98}. Galaxy environment is taken from the
group catalogues of \protect\citet{Garcia93},
\protect\citet{ZabludoffMulchaey98} and
\protect\citet{Whiteetal99}. BGG and BCG stand for Brightest Group Galaxy and 
Brightest Cluster Galaxy, indicating that the galaxy is the brightest (and
presumably dominant) object in the surrounding system. $^{\dag}$ indicates
galaxies which are not identified as brightest group galaxies in the group
catalogues, but lie at the centre of the X-ray halo of their group.}
\end{table*}

\section{Data Reduction}
\label{datared}
Data reduction and analysis of the X--ray datasets were carried out using
the {\sc asterix} software package. Before the datasets could be used,
various sources of contamination had to be removed. Possible sources include
charged particles and solar X--rays scattered into the telescope from the
Earth's atmosphere. Onboard instrumentation provides information which allows
periods of high background to be identified. The master veto counter
records the charged 
particle flux, and we excluded all time periods during which the
master veto rate exceeded 170 count s$^{-1}$. Solar contamination
causes a significant overall increase in the X--ray event rate. To remove
this contamination we excluded all times during which the event rate
deviated from the mean by more than 2$\sigma$. This generally
removed no more than a few percent of each dataset.

After this cleaning process each dataset was binned into a
3--dimensional (x, y, energy) data cube. Spectra or images can be
extracted from such a cube by collapsing it along the axes.  A model
of the background was then generated based on an annulus taken from
this cube.  We used annuli of width 0.1$\deg$, and inner radius
0.4$\deg$ where possible. In cases where this would place the annulus
close to the source we moved the annulus, generally to \mbox{r =
0.55$\deg$}. To ensure that the background model was not biased by
sources within the annulus, an iterative process was used to remove
point sources of $> 4.5~\sigma$ significance. A number of our galaxies
are found within groups and clusters of galaxies, many of which have
their own X--ray halos. Our intention was to model these spectrally
and spatially in order to accurately remove the effects of their
contamination of our target galaxies. We therefore moved the annulus
outward to avoid the emission, where possible. In cases where the
emission appeared to extend to the edge of the field of view, we used
a background annulus at r=0.9$\deg$. This occurred for a small number
of galaxies which lie in the centres of clusters (\eg NGC~1399). The
use of a background annulus which lies within the cluster emission
means that we are likely to overestimate the true background and hence
over-correct for it. However, as we are using the largest annulus
possible, we should mimimize the degree of overestimation. We can also
expect the central galaxy component of the emission to have a much
higher surface brightness than the cluster emission, so that
oversubtraction will have a negligible effect on it. Surface
brightness fits should therefore be accurate for the central
component, which is our main interest, and as good as is possible for
the cluster component.

The resulting background model was then used to produce a
background-subtracted cube. Regions near the {\it PSPC} window support
structure were
removed from these images, as objects in those areas would have been
partially obscured during the observation. The cube was further corrected
for dead time and vignetting effects, and point sources were removed. 

Examination of background subtracted images allowed us to locate each
galaxy and produce a radial profile of the surrounding region. From these
profiles, regions of interest were selected, from which images or spectra
for use in fitting could then be extracted. The majority of our galaxies
are known to be members of groups or clusters, and as such we expected to
see emission from an intergalactic medium (IGM) surrounding them. In cases
where the galaxy did not appear to be contaminated with other emission, we
defined the region of interest (RoI) as being within the radius at which
the emission dropped to the background level. This region was suitable for
both spatial and spectral fitting. In cases where contaminating
intergalactic emission was seen, we defined separate regions of interest,
one for spectral and one for spatial fits.  For spatial fitting, we again
define the RoI as being within the radius at which emission drops to the
background level. This will contain both the galaxy and the surrounding
group or cluster, allowing us to fit models to both and thereby accurately
remove contaminating emission.  A background annulus for use with this
region was selected as described above.  

For spectral fitting in cases where the galaxy is surrounded by
contaminating group/cluster emission, a smaller RoI was defined, using the
radius at which the galaxy emission dropped to the level of the surrounding
group or cluster halo.  Within this radius, emission should be dominated by
components associated with the galaxy, though it may still be contaminated
by group/cluster emission along the line of sight. The emission outside
this radius should be primarily produced by the group or cluster halo. We
therefore take a local background spectrum from an annulus with an inner
radius 0.05$\deg$ larger than the new region of interest, and use this to
generate a local background model. This local background should account for
both the cosmic X--ray background in the region of interest, and for the
group/cluster contamination along the line of sight. Depending on the form
of the group/cluster halo, we might expect to under-subtract this
contamination to some degree. For example, if the cluster halo is steeply
declining in surface brightness outside our region of interest, the local
background annulus will contain fewer counts and we would expect to
underestimate the contamination along the line of sight. Ideally we would
hope that any extended group or cluster halo would have a core radius
somewhat larger than the region of interest, so that its surface brightness
is relatively constant over the whole area we are considering. Any serious
under-subtraction of group or cluster component will have an effect on the
spectral fits we obtain, particularly in the more massive clusters where
contamination by the surrounding hot ICM would produce fits with higher
than expected temperatures. Similarly, if we have misjudged the radius at
which emission associated with the galaxy becomes less important than that
associated with the surrounding structure, we might expect to subtract part
of the galaxy emission. Again, we would expect to see evidence of this in
the results of spectral fits to the data. We return to this question in Section~\ref{sec:results}.

\section{Spectral and Spatial Analysis}
\label{analysis}
Spectra for each galaxy were obtained by removing all data outside the
region of interest and collapsing the data cube along its x and y axes. 
As with the background annulus, an iterative process was used to remove
point sources of $>$4.5 $\sigma$ significance in the region of interest,
although any point sources within the D$_{25}$ diameter were assumed to be
associated with the galaxy itself and therefore not removed. The 
spectra could then be fitted with a variety of models. To provide a
baseline for later fits and to measure the basic properties of the galaxy
halo, each 
spectrum was fitted with a MEKAL hot plasma model (\citealt{Kaastramewe93};
\citealt{Liedahletal95}). Initially, only normalisation was fitted. Hydrogen
absorption column densities were fixed at values determined from radio
surveys (\citealt{Starketal92}), and temperature and metal abundance
were fixed at 1 keV and 1 solar respectively. Parameters were then freed in 
order (temperature, hydrogen column, metallicity), and only re-frozen at
their starting values if they became poorly defined or tended to extreme
values. The basic temperature and metallicity values are likely to be
representative of the majority of early--type galaxies
(\citealt{Matsushita00}; \citealt{Matsushita00b}), but clearly fitted values are 
preferable.

We then attempted to fit two component spectral models for each
galaxy. These generally included a power--law + MEKAL model and
bremsstrahlung + MEKAL model in which the bremsstrahlung temperature was
fixed at 7 keV. The first component was intended to represent
a hard component produced by the
population of X--ray binaries and other unresolved stellar sources within each
galaxy. Recent {\it Chandra} studies of extragalactic X--ray binary
populations suggest that this emission is reasonably modeled by a
power--law of index $\sim$1.2 (\citealt{SarazinIrBreg01};
\citealt{Blantonetal01}), while {\it ASCA} studies show good fits using a
high temperature bremsstrahlung model \citep{Matsushita00}. 
All models were fitted using the Cash statistic
\citep{Cash79}. The Cash statistic is defined as -2ln$L$ where $L$ is the
likelihood function. This means that the most likely model has a minimum
Cash statistic and that differences in the statistic are chi-squared
($\chi^2$) distributed. Thus confidence intervals can be calculated in the
same was as for a conventional $\chi^2$ fit.
By comparing the best fit Cash statistic for each model,
and visually examining the spectral fit, we selected the best fit model for
each galaxy. From this we could extract (in most cases) the X--ray temperature
and metallicity of the galaxy halo, as well as the X--ray flux from the
galaxy halo and the stellar contribution.

For each galaxy in the sample, we also derived simple projected temperature 
and hardness profiles. Temperature profiles were produced by splitting the
larger, surface brightness region of interest into several annuli, from
which spectra were extracted. These spectra were then fitted using the best 
fitting spectral model for the galaxy as a whole. Initially the models were 
fitted with the metallicity and hydrogen column density frozen at their
global best fit values. However, if the data quality permitted, we freed
these parameters, providing us with crude metallicity profiles for a
fraction of our sample. Given the limited spectral range of {\it ROSAT}, 
the inability of the PSPC to resolve individual spectral lines, and the
small number of counts in each annulus, the abundances fitted should not be 
taken as accurate measurements. However, in some cases they do show
interesting trends when considered in conjunction with the temperature
profiles. 

Hardness profiles were calculated in a somewhat similar manner. Again the
larger region of interest was split in to a number of annuli. From each of
these, counts in soft (0.3--1.3 keV) and hard (1.3--2.4 keV) bands were
extracted and divided to produce a ratio of hard/soft emission. Simulated
spectra indicate that a 0.5 keV MEKAL spectrum produces a value of $\sim$
0.5, while a power law of $\Gamma$=1.7 and a 7 keV bremsstrahlung spectrum
produce values of $\sim$1.1 and 1.2 respectively. These profiles can be
used to give a basic idea of changes in emission across the galaxy and in
particular to identify AGN.

In order to study the spatial properties of the galaxy X--ray emission, we 
also performed fits to the 2--dimensional surface brightness profile of each
galaxy. Following the initial data reduction described in
Section~\ref{datared}, we extracted an image in the 0.5--2 keV band and
corrected it for vignetting. This was done using an energy--dependent
exposure map (see \citealt{Snowdenetal94} for a full description). Point
sources were removed as in the spectral analysis, and unrelated extended
sources identified and excluded by hand. Use of the energy dependent
exposure map results in a constant background level across the image,
so a flat background was also determined and subtracted from the data.

As in the case of spectral analysis, we can choose to fit a variety of
surface brightness models to our data. The most commonly used in this work
was a modified King function (or ``$\beta$--profile'') of the form:

\begin{equation}
  S(r) = S_0(1+(r/r_{core})^2)^{-3\beta_{fit}+0.5}
\label{eqn:kingSB}
\end{equation}

where $S(r)$ is the surface brightness at a given radius, $S_0$ is the
central surface brightness $r_{core}$ is the core radius and \Bfit\
is a measure of the slope of the surface brightness profile. At various
stages of the analysis we also fitted point source models and  
de Vaucouleurs $r^{1/4}$ law models, using the form:

\begin{equation}
  S(r) = S_e \cdot {\rm exp}\{-7.67[(r/r_e)^{0.25}-1]\}
\end{equation}

where $S_e$ is the surface brightness at $r_e$, the effective radius (the
isophotal radius containing half the total luminosity.)

\begin{figure*}
\centerline{\epsfig{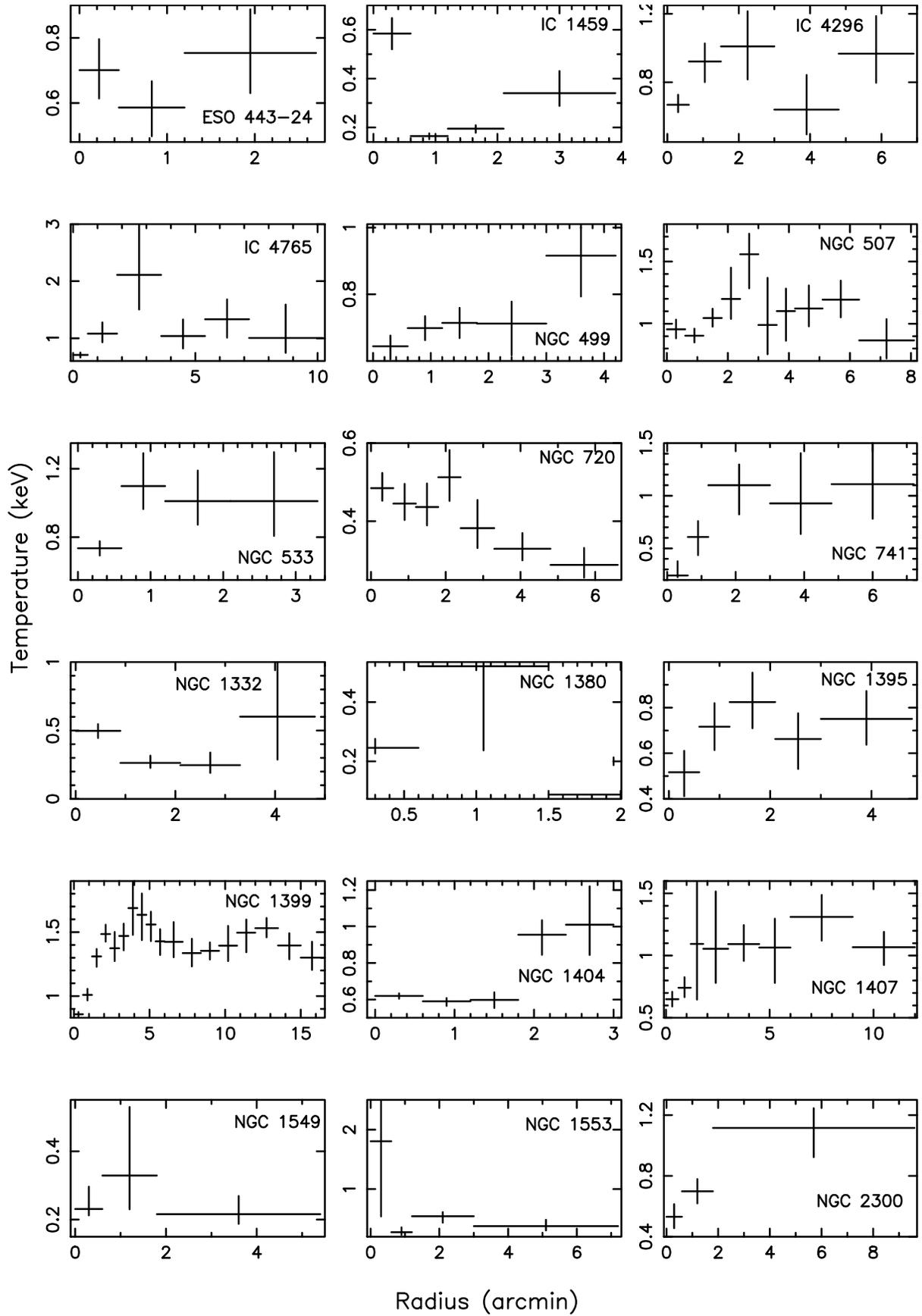}}
\vspace{-1cm}
\caption{\label{fig:Tprof}Measured projected temperature profiles for our
  sample of galaxies} 
\end{figure*}
\begin{figure*}
\centerline{\epsfig{file=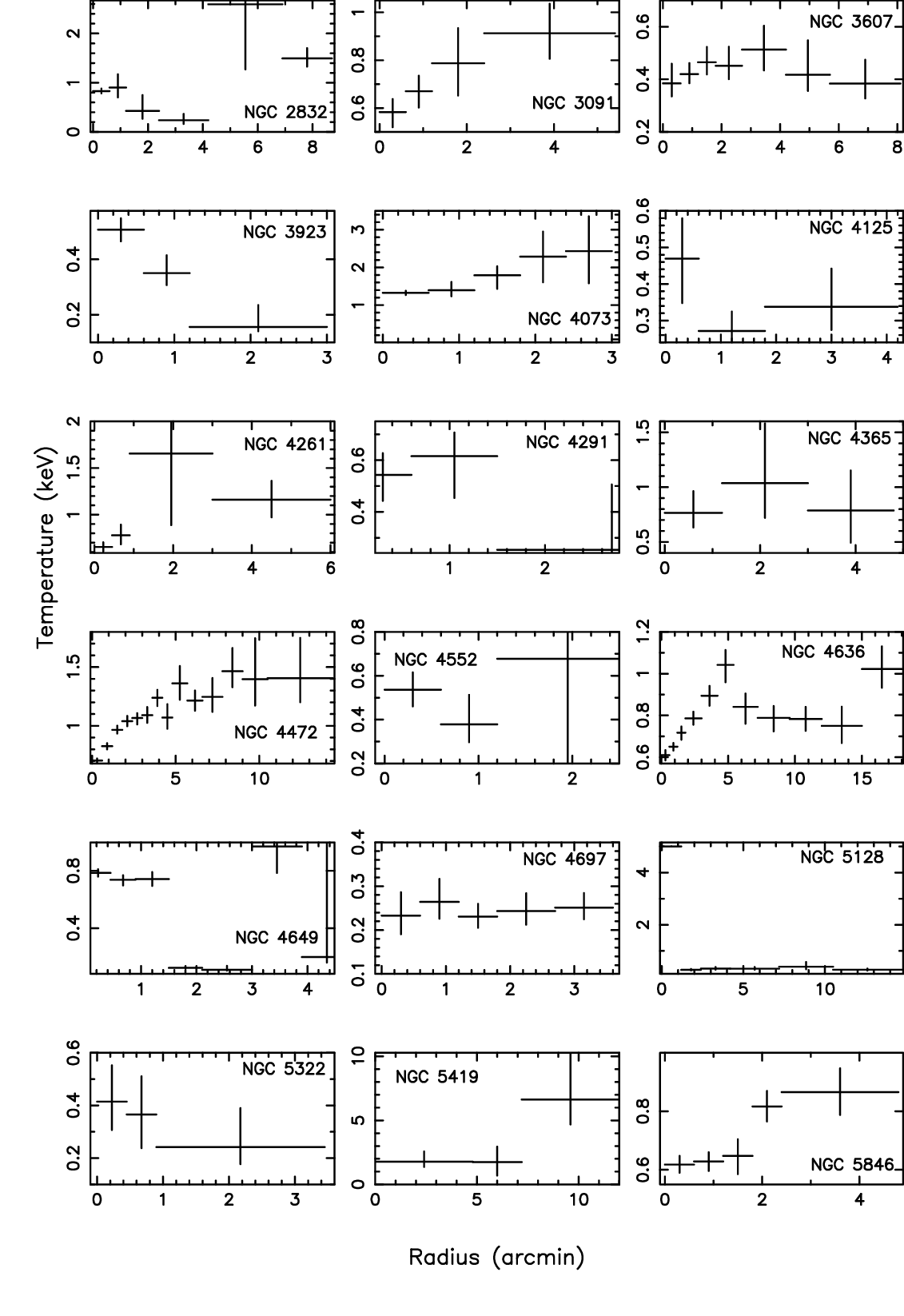,width=6.5in}}
\vspace{-1cm}
\contcaption{}
\end{figure*}
\begin{figure*}
\centerline{\epsfig{file=T_prof_3.eps,width=6.5in}}
\vspace{-19cm}
\contcaption{}
\end{figure*}

\begin{figure*}
\vspace{-2cm}
\centerline{\epsfig{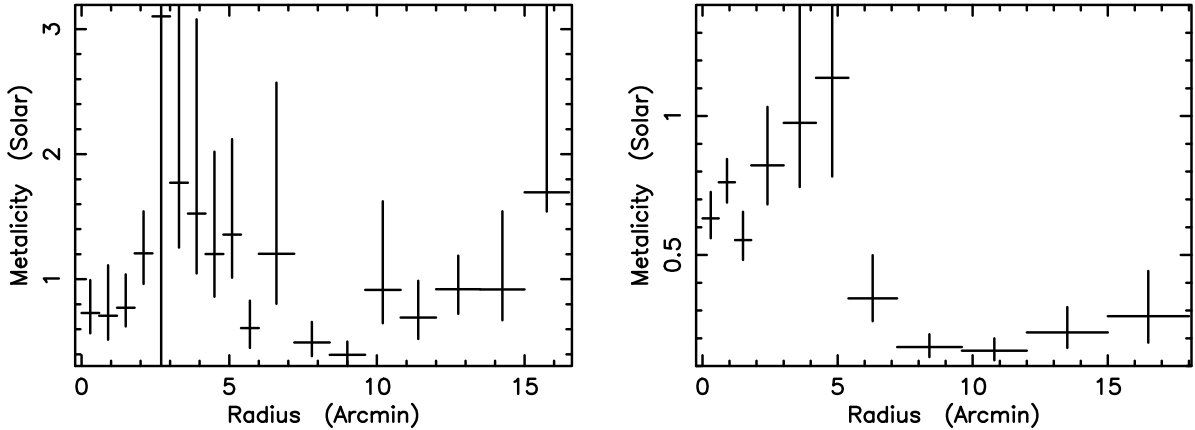}}
\vspace{-15.5cm}
\caption{\label{fig:Zprofs} Fitted metallicity profiles of NGC~1399 (left
  panel) and NGC~4636 (right panel)}
\end{figure*}

Models were convolved with the PSPC point spread function at an energy
determined from the mean photon energy of the emission in the region of
interest and then fitted to the data. Both spherical and elliptical fits
were possible when using the King and de Vaucouleurs models, with the
position angle and major to minor axis ratio measuring the shape and
orientation of elliptical fits. When using King models, all parameters
(core radius, \Bfit\, normalisation, x and y position and the ellipticity
parameters) were usually allowed to vary freely, as were the parameters in
point source models (x and y position, normalisation). The de Vaucouleurs
model is intended to represent the unresolved discrete source population of
the galaxies, which we assume will take the same form as the stellar
population. An alternative approach would be to assume that the discrete
source population follows the distribution of globular clusters, but we
lack accurate spatial models of this distribution for many of our galaxies.
We therefore initially set the effective radius parameter of the de
Vaucouleurs models to the value of the optical effective radius, and held
it frozen in most cases. We did allow the effective radius to vary for a
small number of galaxies, where the fit was well constrained by the data,
in order to investigate differences between the optical and X--ray stellar
profiles. However only one galaxy, NGC~4697, was best fit by a model
including a de Vaucouleurs component, demonstrating that in most of the
galaxies in the sample, either emission from hot gas dominates or the X-ray
binary population does not follow the stellar population. A \chandra\ 
observation of NGC~4697 has shown it to have a relatively small gas halo,
with much of its emission contributed by point sources
\citep{SarazinIrBreg01}, so the success of the de Vaucouleurs component in
this case could is perhaps unsurprising. The \chandra\ data for this galaxy
are best fit by a surface brightness model which includes a de Vaucouleurs
component and a King model whose parameters are such that it is flat,
providing a fairly constant contribution over the area studied.

The use of 2--dimensional datasets to fit the surface brightness
distribution can result in a low number of counts in many of the data
bins. Under these conditions $\chi^2$ fitting performs poorly
\citep{NousekShue89} so, as in the spectral analysis, maximum likelihood
fitting based on the Cash statistic was used. However, the Cash statistic
gives no indication of the absolute quality of the fit, only the quality
relative to other fits. In order to gain some estimate of the true fit
quality, we used a Monte Carlo approach, in which the best fit 1- and
2-component model was used to generate 1000 images of the groups, to which
Poisson noise was added. These were then compared to the original image,
the Cash statistic determined, and a Gaussian fitted to the resulting
spread of values. By comparing the actual Cash statistic to this
distribution of values, we were able to determine the probability that the
model could have produced the data. We were therefore able to identify
cases where the 2-component fit was no more likely to reproduce the data than
the 1-component fit, and discard the 2-component fits for these galaxies.

\section{Results}
\label{sec:results}

\subsection{Spectral and spatial fits}

Table~\ref{tab:spec} shows the results of our spectral fits. As
mentioned previously, metal abundances from \rosat\ PSPC spectra are
inherently unreliable, due to the relatively poor spectral resolution
of the instrument. This is reflected by the large errors on some of
our fitted values, and by the fact that in some cases we had to hold
metallicity frozen in order to secure a stable fit. The temperature
values are more reliable, and give a mean temperature of 0.67$\pm$0.29
keV.  

As discussed in Section~\ref{datared}, a poor choice of local background
for our targets could result in spectral fits biased by
inclusion of group or cluster emission, or accidental subtraction of
some of the galaxy emission. The clearest sign of this bias would be
unusually high or low fitted temperature, significantly different from
those found in other studies. Four of our galaxies have temperatures above
1 keV, and one (NGC 4697) has a temperature lower than 0.3 keV. These are
outside the range commonly considered typical for elliptical galaxies, so
we compare the results for these galaxies to those in the literature. NGC
507 has a temperature marginally above 1 keV. Previous \rosat\ and \asca\ 
studies have found similar temperatures
\citep{KimFabbiano95,Matsumotoetal97} and metal abundances \citep{Buote00},
and more recent \chandra\ data also supports a temperature of $\sim$1 keV
\citep{Formanetal01}. \citet{Buote02} fits a two temperature model to \xmm\ 
EPIC data for NGC 1399, and recovers temperatures of $\sim$1.5 and
$\sim$0.9 keV within 1\arcm, with both components approaching a temperature
of 1.3-1.5 keV at 3-10\arcm. Our value of $\sim$1.2 keV is quite comparable
to the cooler component, considering the region from which our spectrum was
extracted. NGC 4073 has the most extreme temperature in
Table~\ref{tab:spec}, kT=1.6 keV. Analysis of \xmm\ EPIC data for NGC 4073
and its surrounding group \citep[][in prep.]{O'Sullivanetal03} suggest a
temperature gradient within the stellar body of the galaxy, with projected
temperatures rising from 1.4 to $\sim$2 keV. We also find a high
temperature for NGC 6269, kT=1.4$\pm$0.2 keV. No \xmm\ or \chandra\ results
are available in the literature for NGC 6269, but previous analyses of the
\rosat\ data have found temperatures of 1.3$\pm$0.15 keV
\citep{DahlemThiering00} and 1.36$\pm$0.07 keV \citep{Mulchaeyetal96}.
These are identical within the errors with our result. Lastly, NGC 4697 was
one of the first elliptical galaxies to be observed with \chandra, which
showed it to possess a relatively cool gas component with kT$\sim$0.29 keV
\citep{SarazinIrBreg01}. This is a fairly good match to our measured
temperature of kT=0.24$\pm$0.2 keV, and it should be noted that at least
part of the difference between these results may arise from \chandra\ 
calibration issues. In general, these comparisons suggest that our method
of background selection and data analysis produces fairly accurate fits to
the data.

When calculating the luminosities of our targets, we are faced with the
difficulty that while we have surface brightness models which should
provide an accurate estimate of the total number of counts from the galaxy
halo, we only have spectral information for a smaller central region.
Ideally we would be able to simultaneously fit spectral and spatial data,
giving a true luminosity for each component
\citep[\textit{e.g.}][]{Lloyd-Daviesetal00}. In practice we have chosen to
scale up the MEKAL component of the spectral fit to the number of counts
found for the surface brightness model. This allows us to
calculate a gas luminosity for the galaxy component, but ignores the
contribution from discrete sources. This luminosity will therefore be an
overestimate of the true gas luminosity associated with the galaxy.
However, we expect large optically luminous galaxies such as our targets to
be almost entirely dominated by gas emission. With a small number of
exceptions, the spectral fits confirm this, suggesting that in most cases
the overestimation is small. It is also notable that because the
bremsstrahlung component peaks at a higher energy than the MEKAL, a given
luminosity corresponds to a smaller number of bremsstrahlung counts (in the
\rosat\ band)than it would for a MEKAL model. This means that our
overestimate of luminosity is reduced, as the number of counts associated
with the bremsstrahlung component, and assumed in the scaling to be part of
the MEKAL component, will produce only a small increase in gas luminosity.

\begin{table*}
\begin{tabular}{lllllcc} \hline
Name & \NH\ & \Tx\ & $Z$ & Log \Lx & Model & Profile \\
  & ($\times$10$^{-21}$ cm$^{-2}$) & (keV) & (\Zsol) & (erg s$^{-1}$) & & \\
\hline\\[-3mm]
ESO~443-24 &$0.50^{+5.3}_{-0.19}$  &$0.69^{+0.07}_{-0.39}$  &$1.0^{+1.0}_{-0.7}$      &$41.78^{+1.}_{-0.06}$ &MK+BR & I \\
IC~1459    &$0.069\pm 0.012$       &$0.51\pm 0.05$          &$1.0$             &$40.3\pm 0.2$         &MK+BR & H \\
IC~4296    &$0.63^{+0.33}_{-0.24}$ &$0.72\pm 0.07$          &$0.23\pm 0.92$           &$41.7^{+0.3}_{-0.8}$  &MK+BR & C \\
IC~4765    &$2.6^{+3.8}_{-0.9}$    &$0.64^{+0.09}_{-0.37}$  &$0.17\pm 0.17$           &$42.4^{+0.7}_{-0.1}$  &MK+BR & C \\
NGC~499    &$0.79^{+0.32}_{-0.14}$ &$0.70^{+0.02}_{-0.03}$  &$1.0$             &$42.6\pm 0.3$         &MK+BR & I \\
NGC~507    &$0.59\pm 0.09$         &$1.03\pm 0.05$          &$0.95^{+2.2}_{-0.33}$    &$42.9^{+0.4}_{-0.7}$  &MK+BR & C \\
NGC~533    &$0.29\pm 0.03$         &$0.84^{+0.05}_{-0.04}$  &$1.1\pm 1.1$             &$42.2\pm 0.2$         &MK+BR & C \\
NGC~720    &$0.15\pm 0.15$         &$0.50\pm 0.04$          &$0.28^{+0.06}_{-0.04}$   &$41.1\pm 0.3$         &MK+BR & H \\
NGC~741    &$0.54^{+0.11}_{-0.08}$ &$0.71\pm 0.07$          &$0.26\pm 0.26$           &$41.9^{+0.3}_{-0.4}$  &MK+BR & C \\
NGC~1332   &$0.16^{+0.04}_{-0.03}$ &$0.41^{+0.06}_{-0.05}$  &$0.59\pm 0.59$           &$40.6\pm 0.2$         &MK+BR & H \\
NGC~1380   &$0.18\pm 0.10$         &$0.30^{+0.08}_{-0.05}$  &$0.11^{+0.44}_{-0.07}$   &$40.4^{+0.4}_{-1.0}$  &MK+BR & ? \\
NGC~1395   &$0.069\pm 0.016$       &$0.65^{+0.04}_{-0.05}$  &$1.0$             &$40.6\pm 0.2$         &MK+BR & C \\
NGC~1399   &$0.12\pm 0.01$         &$1.21\pm 0.03$          &$1.1^{+0.3}_{-0.2}$      &$41.8\pm 0.3$         &MK+BR & C \\
NGC~1404   &$0.18\pm 0.02$         &$0.60\pm 0.01$          &$0.35^{+0.05}_{-0.04}$   &$41.66^{+0.04}_{-0.05}$ &MK & C \\
NGC~1407   &$0.72^{+0.22}_{-0.16}$ &$0.79^{+0.08}_{-0.07}$  &$0.14^{+0.14}_{-0.06}$   &$41.5^{+0.2}_{-0.3}$  &MK+BR & C \\
NGC~1549   &$0.046\pm 0.046$       &$0.25^{+0.08}_{-0.05}$  &$0.14^{+0.27}_{-0.08}$   &$39.7^{+0.2}_{-0.4}$  &MK+BR & I \\
NGC~1553   &$0.045\pm 0.033$       &$0.53\pm 0.15$          &$0.10^{+0.16}_{-0.05}$   &$40.2^{+0.3}_{-0.4}$  &MK+BR & ? \\
NGC~2300   &$0.97^{+0.97}_{-0.30}$ &$0.62^{+0.07}_{-0.14}$  &$0.23\pm 0.23$           &$41.43^{+0.2}_{-0.07}$ &MK+BR & ? \\
NGC~2832   &$0.13\pm 0.02$         &$0.82\pm 0.05$          &$1.0$             &$41.9\pm 0.2$         &MK+BR & ? \\
NGC~3091   &$0.38^{+0.08}_{-0.06}$ &$0.64^{+0.04}_{-0.05}$  &$1.0$             &$41.7\pm 0.2$         &MK+BR & C \\
NGC~3607   &$0.015\pm 0.015$       &$0.45\pm 0.06$          &$0.71^{+0.57}_{-0.21}$   &$40.7^{+0.3}_{-0.4}$  &MK+BR & I \\
NGC~3923   &$0.58\pm 0.22$         &$0.46^{+0.04}_{-0.05}$  &$1.0$             &$40.8\pm 0.2$         &MK+BR & H \\
NGC~4073   &$0.16^{+0.02}_{-0.03}$ &$1.6\pm 0.2$            &$1.6^{+1.1}_{-0.4}$      &$43.1^{+0.1}_{-0.3}$  &MK+BR & C \\
NGC~4125   &$0.10\pm 0.04$         &$0.34^{+0.05}_{-0.04}$  &$0.34\pm 0.34$           &$41.27\pm 0.07$       &MK & ? \\
NGC~4261   &$0.087\pm 0.017$       &$0.67^{+0.04}_{-0.05}$  &$1.3\pm 1.3$             &$41.0\pm 0.2$         &MK+BR & C \\
NGC~4291   &$0.22^{+0.18}_{-0.09}$ &$0.59^{+0.06}_{-0.07}$  &$0.63^{+4.8}_{-0.13}$    &$41.20^{+0.4}_{-0.02}$ &MK & I \\
NGC~4365   &$0.11^{+0.06}_{-0.05}$ &$1.0^{+0.3}_{-0.2}$     &$0.064^{+0.11}_{-0.053}$ &$40.5\pm 0.1$         &MK & I \\
NGC~4472   &$0.162\pm 0.007$       &$0.88\pm 0.01$          &$1.0$             &$41.5\pm 0.2$         &MK+BR & C \\
NGC~4552   &$0.18\pm 0.03$         &$0.54\pm 0.06$          &$1.0$             &$40.7\pm 0.2$         &MK+BR & ? \\
NGC~4636   &$0.25^{+0.05}_{-0.06}$ &$0.55\pm 0.03$          &$0.40^{+0.32}_{-0.13}$   &$42.0^{+0.1}_{-0.2}$  &MK+BR & C \\
NGC~4649   &$0.24^{+0.06}_{-0.07}$ &$0.78\pm 0.02$          &$0.80^{+1.0}_{-0.36}$    &$39.3^{+0.3}_{-0.5}$  &MK+BR & C \\
NGC~4697   &$0.16\pm 0.03$         &$0.24\pm 0.02$          &$0.40\pm 0.40$           &$39.0\pm 0.2$         &MK+BR & I \\
NGC~5128   &$0.43^{+0.07}_{-0.05}$ &$0.35^{+0.04}_{-0.03}$  &$1.0$             &$40.2\pm 0.2$         &MK+BR & H \\
NGC~5322   &$0.045\pm 0.037$       &$0.33^{+0.100}_{-0.06}$ &$0.19\pm 0.19$           &$40.3\pm 0.3$         &MK+BR & ? \\
NGC~5419   &$0.29^{+0.07}_{-0.05}$ &$0.69\pm 0.26$          &$0.32\pm 0.32$           &$42.0^{+0.3}_{-0.4}$  &MK+BR & ? \\
NGC~5846   &$0.30\pm 0.03$         &$0.66\pm 0.02$          &$1.0$             &$40.5^{+0.2}_{-0.3}$  &MK+BR & C \\
NGC~6269   &$0.53\pm 0.12$         &$1.4\pm 0.2$            &$0.33^{+0.36}_{-0.17}$   &$43.4\pm 0.1$         &MK & C \\
NGC~6482   &$0.68^{+0.72}_{-0.19}$ &$0.55^{+0.04}_{-0.07}$  &$1.0^{+1.0}_{-0.5}$      &$41.2^{+0.5}_{-0.4}$  &MK+BR & I \\
NGC~7619   &$0.34^{+0.08}_{-0.06}$ &$0.81\pm 0.03$          &$1.9\pm 1.9$             &$42.0^{+0.3}_{-0.8}$  &MK+BR & C \\
\hline
\end{tabular}
\caption{\label{tab:spec} Results of the spectral fits to our sample
  galaxies. Where possible, a absorbed MEKAL+bremsstrahlung (MK+BR) model
  was fitted, but in cases where the bremsstrahlung normalisation always
  tended to zero, this component was removed from the fit. All upper and
  lower (1$\sigma$) errors on fitted parameters were calculated individually, but are
  shown as a single $\pm$ error when they are identical to two significant
  figures. Those galaxies for which metallicity could not be successfully
  fitted are listed with a fixed solar metallicity with no
  errors. Temperature profiles are classified as isothermal (I), cool core
  (C), hot core (H), or uncertain (?).}
\end{table*}

\begin{figure}
\centerline{\epsfig{file=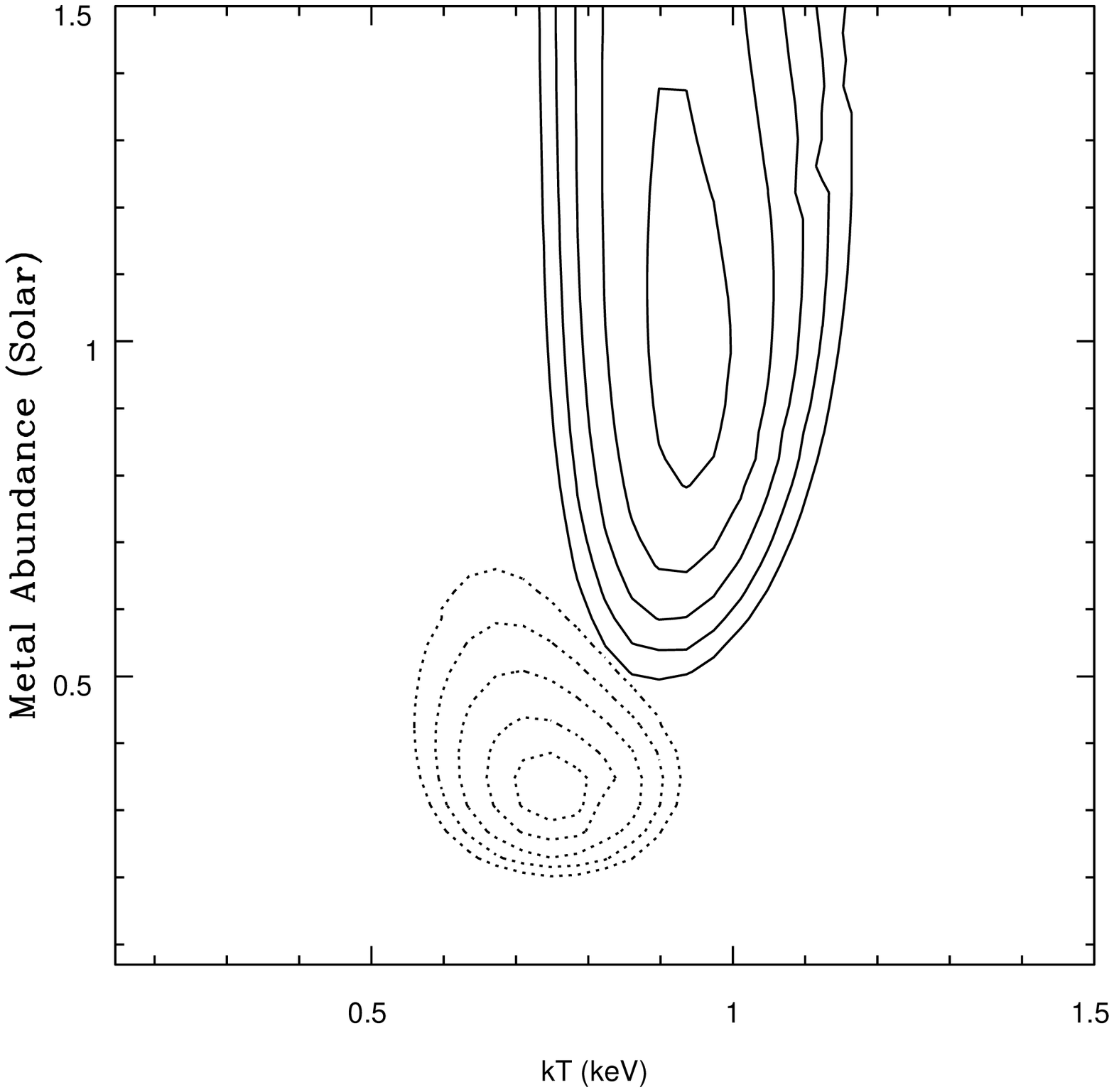, width=9cm}}
\vspace{-3cm}
\caption{\label{fig:conreg} Confidence regions for bins 6 (solid lines)
  and 7 (dotted lines) of the temperature and metallicity profiles of
  NGC~4636. The contours show confidence levels of 1, 3, 6, 9 and
  12$\sigma$, and only at the 12$\sigma$ level do the two regions come
  close to overlap.}
\end{figure}

Temperature profiles for our sample of galaxies are shown in
figure~\ref{fig:Tprof}. The extent of the profiles varies, owing to the
relative quality of data, length of exposure and galaxy distance.
Comparison of these profiles with those shown in \citet{Helsdonponman00}
shows them to be similar in most cases where the samples overlap. We have
classified the galaxy profiles into four groups; cooling cores (\eg
NGC~1399), hot cores (\eg NGC~720), isothermal (\eg NGC~4697) and those
where the data quality prevents a judgment (\eg NGC~4552, where the errors
on the outermost bin are large enough to make it suspect). Cool and hot
core galaxies are selected under the requirement that their central bin
must be hotter or cooler than an average outer temperature by at least
20\%, and that the errors in \Tx\ must be smaller than this amount. It is
also possible to see evidence of AGN activity in some of the profiles,
particularly in the case of NGC~5128, where the central bin has a
temperature of $\sim$5 keV and the rest of the galaxy $<$1 keV. An
interesting feature of some of the better defined profiles which show
central cooling (\eg NGC~507, NGC~1399, NGC~4636) is that the temperature
rises with radius to a value above the apparent outer mean temperature, and
the falls back to that mean, producing a temperature peak at moderate
radii. As all of these galaxies are embedded in larger group or cluster
halos, this peak may mark the boundary of a group or cluster scale cooling
flow, or the point of interaction between the galaxy and its environment.
The observations available for NGC~1399 and NGC~4636 contain very large
numbers of counts, allowing us to include metallicity as a free parameter
in the profile fitting. Metallicity profiles of these two galaxies are
shown in Figure~\ref{fig:Zprofs}.  Despite the large errors in some bins,
it is notable that both metallicity profiles follow the same structure as
seen in temperature; a central trough, rising to a peak at moderate radius,
with an outer region of relatively low abundance. The peak temperature and
metallicity occur at approximately the same radius in both cases. In order
to check that correlations between temperature and abundance in the fits
were not biasing the results we modeled the fit space for the two bins on
either side of the apparent break in the profile of NGC~4636, calculating
fit statistics at a range of temperatures and metallicities. Comparing the
confidence regions for the two points shows that they are dissimilar to at
least 9$\sigma$ significance, strongly suggesting the break in the profile
is real. Confidence regions for the two bins are plotted in
Figure~\ref{fig:conreg}.

\begin{table*}
\begin{tabular}{lllllcc} \hline
Name &Core Radius &\Bfit\ &Axis Ratio &Position Angle &RoI Radius &Model  \\
  & (arcmin) &  &  & (degrees) & (degrees) &   \\ \hline\\[-3mm]
ESO~443-24 &$0.106\pm 0.004$ &$0.55\pm 0.03$ &$1.7^{+0.3}_{-0.2}$ &$306\pm 7$ &$0.045$ &KI \\
IC~1459 &$0.17\pm 0.02$ &$0.89^{+0.36}_{-0.09}$ &$1.6^{+1.4}_{-0.2}$ &$276\pm 7$ &$0.065$ &KI+KI \\
IC~4296 &$0.080\pm 0.004$ &$0.66^{+0.18}_{-0.02}$ &$2.2^{+0.8}_{-0.4}$ &$29^{+10}_{-8}$ &$0.12$ &KI+KI \\
IC~4765 &$0.39\pm 0.02$ &$1.24^{+0.6}_{-0.06}$ &$1.1^{+0.1}_{-0.7}$ &$190$ &$0.17$ &KI+KI \\
NGC~499 &$0.85\pm 0.04$ &$0.70\pm 0.03$ &$1.05^{+0.03}_{-0.04}$ &$290^{+30}_{-20}$ &$0.25$ &KI+KI \\
NGC~507 &$0.82\pm 0.03$ &$0.54\pm 0.03$ &$1.05^{+0.04}_{-0.8}$ &$270$ &$0.14$ &KI+KI \\
NGC~533 &$0.058\pm 0.002$ &$0.53^{+0.04}_{-0.02}$ &$1.24\pm 0.06$ &$240\pm 10$ &$0.16$ &KI+KI \\
NGC~720 &$0.15\pm 0.02$ &$0.483^{+0.010}_{-0.009}$ &$1.34^{+0.09}_{-0.08}$ &$294^{+20}_{-7}$ &$0.11$ &KI+KI \\
NGC~741 &$0.172\pm 0.006$ &$0.65^{+0.64}_{-0.06}$ &$1.41^{+0.08}_{-0.1}$ &$128^{+10}_{-9}$ &$0.12$ &KI+KI \\
NGC~1332 &$0.010\pm 0.001$ &$0.548^{+0.009}_{-0.008}$ &$1.8^{+0.3}_{-0.2}$ &$295\pm 4$ &$0.080$ &KI+KI \\
NGC~1380 &$0.090\pm 0.011$ &$0.51^{+0.04}_{-0.03}$ &$1.1^{+0.2}_{-0.3}$ &$210$ &$0.045$ &KI \\
NGC~1395 &$0.23\pm 0.03$ &$0.52^{+0.06}_{-0.03}$ &$1.3\pm 0.1$ &$41^{+9}_{-10}$ &$0.10$ &KI+KI \\
NGC~1399 &$0.11\pm 0.01$ &$0.59\pm 0.02$ &$1.23^{+0.04}_{-0.03}$ &$176\pm 3$ &$0.12$ &KI+KI \\
NGC~1404 &$0.32\pm 0.04$ &$0.770^{+0.005}_{-0.004}$ &$1.05^{+0.02}_{-0.01}$ &$294^{+10}_{-6}$ &$0.20$ &KI+KI \\
NGC~1407 &$0.18\pm 0.02$ &$0.56^{+0.02}_{-0.01}$ &$1.20\pm 0.06$ &$91^{+7}_{-8}$ &$0.20$ &KI+KI \\
NGC~1549 &$0.021\pm 0.003$ &$0.509\pm 0.003$ &$1.56^{+0.10}_{-0.09}$ &$174^{+7}_{-8}$ &$0.090$ &KI \\
NGC~1553 &$0.43\pm 0.07$ &$0.66^{+0.06}_{-0.09}$ &$1.4^{+0.2}_{-0.1}$ &$304^{+10}_{-9}$ &$0.17$ &KI+KI \\
NGC~2300 &$0.23\pm 0.02$ &$0.69^{+0.12}_{-0.06}$ &$1.1^{+0.2}_{-0.1}$ &$330$ &$0.16$ &KI+KI \\
NGC~2832 &$0.0100\pm 0.0003$ &$0.314\pm 0.007$ &$1.4\pm 0.1$ &$357\pm 6$ &$0.15$ &KI+PS \\
NGC~3091 &$1.17\pm 0.05$ &$1.60^{+0.03}_{-0.04}$ &$1.31\pm 0.04$ &$325^{+7}_{-6}$ &$0.15$ &KI+KI \\
NGC~3607 &$0.63\pm 0.07$ &$0.48\pm 0.02$ &$1.17^{+0.07}_{-0.06}$ &$350\pm 10$ &$0.14$ &KI \\
NGC~3923 &$0.010\pm 0.001$ &$0.55^{+0.05}_{-0.01}$ &$1.8^{+0.4}_{-0.2}$ &$54^{+6}_{-5}$ &$0.080$ &KI+KI \\
NGC~4073 &$0.072\pm 0.002$ &$0.46^{+0.02}_{-0.01}$ &$1.20^{+0.06}_{-0.05}$ &$265^{+7}_{-8}$ &$0.10$ &KI+KI \\
NGC~4125 &$0.017\pm 0.001$ &$0.48^{+0.01}_{-0.08}$ &$1.629^{+0.3}_{-0.003}$ &$269^{+8}_{-9}$ &$0.070$ &KI+PS \\
NGC~4261 &$0.37\pm 0.03$ &$1.2^{+1.4}_{-0.2}$ &$1.8^{+0.3}_{-0.2}$ &$37\pm 5$ &$0.10$ &KI+KI \\
NGC~4291 &$0.38\pm 0.04$ &$0.57^{+0.04}_{-0.03}$ &$1.3\pm 0.1$ &$106^{+9}_{-10}$ &$0.065$ &KI \\
NGC~4365 &$0.54\pm 0.08$ &$0.60^{+0.04}_{-0.03}$ &$2.0\pm 0.2$ &$31\pm 4$ &$0.090$ &KI \\
NGC~4472 &$0.25\pm 0.04$ &$0.597^{+0.009}_{-0.008}$ &$1.08\pm 0.02$ &$83^{+10}_{-9}$ &$0.24$ &KI+KI \\
NGC~4552 &$0.098\pm 0.014$ &$0.60^{+0.02}_{-0.03}$ &$1.7^{+0.2}_{-0.1}$ &$164\pm 5$ &$0.045$ &KI \\
NGC~4636 &$0.40\pm 0.06$ &$0.535^{+0.007}_{-0.006}$ &$1.02^{+0.02}_{-0.09}$ &$24$ &$0.24$ &KI+KI \\
NGC~4649 &$0.13\pm 0.02$ &$0.567\pm 0.008$ &$1.18\pm 0.03$ &$26\pm 6$ &$0.080$ &KI+PS \\
NGC~4697 &$0.99\pm 0.15$ &$0.46^{+0.13}_{-0.07}$ &$1.8\pm 0.3$ &$28^{+6}_{-12}$ &$0.045$ &KI+DV \\
NGC~5128 &$0.90\pm 0.53$ &$0.55\pm 0.01$ &$2.21\pm 0.07$ &$52.0\pm 0.9$ &$0.25$ &KI+KI \\
NGC~5322 &$0.0100\pm 0.0008$ &$0.49\pm 0.01$ &$1.6^{+0.3}_{-0.2}$ &$57^{+8}_{-9}$ &$0.057$ &KI \\
NGC~5419 &$4.5\pm 0.2$ &$0.50^{+0.18}_{-0.07}$ &$1.44\pm 0.06$ &$44\pm 6$ &$0.20$ &KI+PS \\
NGC~5846 &$1.3\pm 0.1$ &$0.80^{+0.05}_{-0.04}$ &$1.15\pm 0.03$ &$45^{+7}_{-6}$ &$0.080$ &KI+PS \\
NGC~6269 &$1.21\pm 0.02$ &$0.40^{+0.04}_{-0.02}$ &$1.19^{+0.1}_{-0.08}$ &$38\pm 15$ &$0.10$ &KI+PS \\
NGC~6482 &$0.162\pm 0.007$ &$0.524^{+0.009}_{-0.001}$ &$1.13^{+0.07}_{-0.05}$ &$220\pm 10$ &$0.10$ &KI+PS \\
NGC~7619 &$0.031\pm 0.002$ &$0.447^{+0.006}_{-0.005}$ &$1.18^{+0.07}_{-0.06}$ &$308\pm 10$ &$0.20$ &KI+KI \\
\hline
\end{tabular}
\caption{\label{tab:SB} Results of the surface brightness fits to our
  galaxies, for the components associated with the galaxy halo. Best fit
  position angle values without errors are given for those galaxies where
  the angle was essentially unconstrained. All errors are quoted at the
  1$\sigma$.}
\end{table*}

Table~\ref{tab:SB} lists the results of the surface brightness fitting for
the galaxy halos of our target galaxies. In the majority of galaxies we
obtain good quality fits, with relatively small errors on the core radius
and slope. As we are able to fit elliptical models, we also list the
position angle of the major axis and axis ratio of the model fits. For five
galaxies, we were unable to determine a reliable position angle, as the
model was consistent (within errors) with being spherical. The best fit
position angles of these galaxies are listed without errors, and should
only be considered as rough estimates.

\subsection{The \LT\ relation}
\label{sec:LT}

\begin{figure*}
\centerline{\epsfig{file=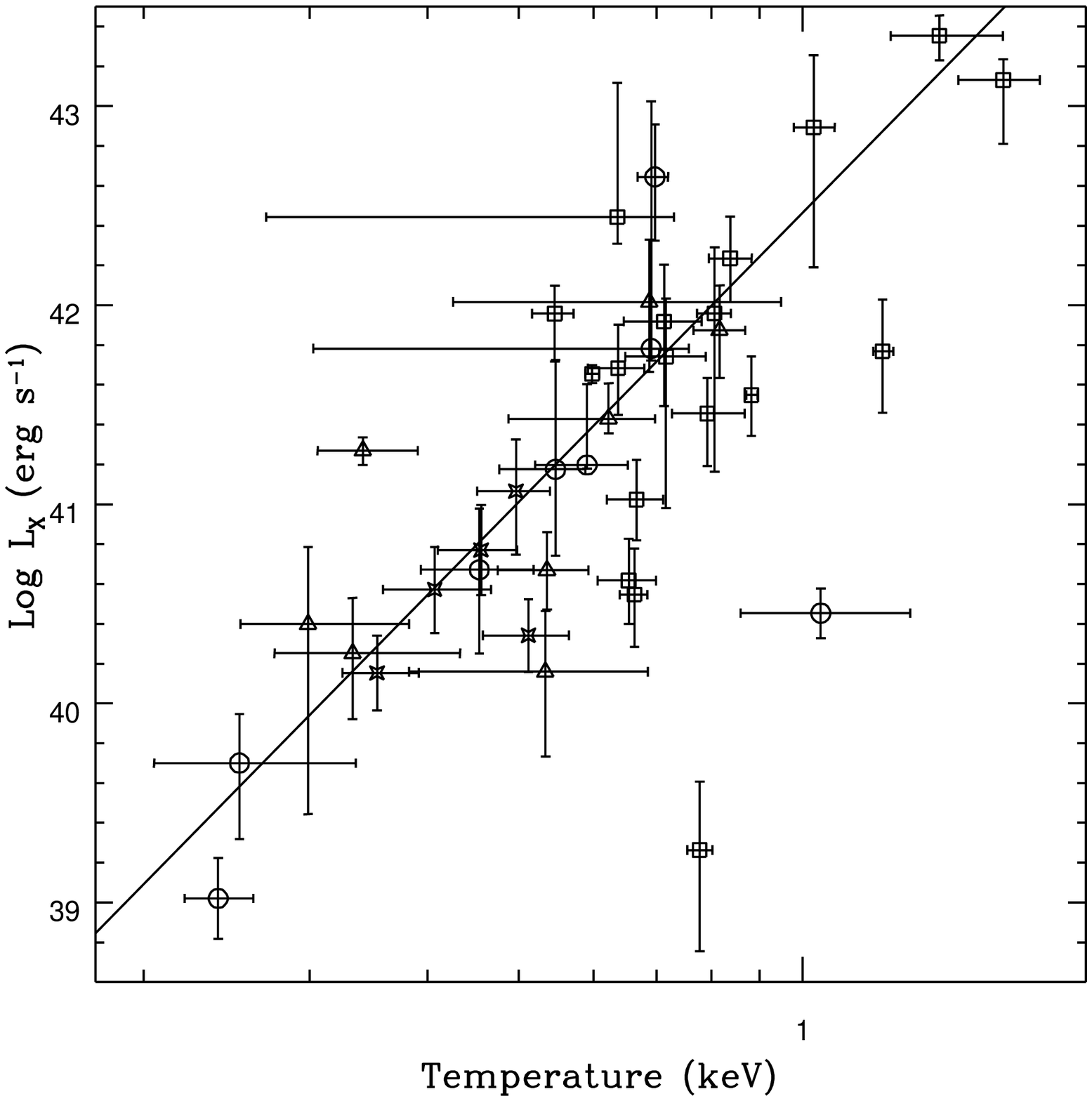,width=6in}}
\vspace{-5cm}
\caption{\label{fig:LTplot} Log \Lx\ plotted against temperature for our
  sample of galaxies. Galaxies which show signs of central cooling
  are marked by open squares, those with central heating by  stars,
  those which are apparently isothermal 
  with  circles, and those which are unclassified by 
  triangles. The solid line shows the best fit line for the complete 
  sample}  
\end{figure*}

Figure~\ref{fig:LTplot} shows \Lx\ plotted against temperature for our
sample.  As can be seen, we find a fairly tight relation between \Lx\ and
\Tx, with only a small number of outlier points. Using Kendall's K-statistic
\citep{Ponmanetal82} to measure the strength of correlation, we find a
significance of $\sim$4.6$\sigma$. We fitted the relation using \textsc{ODRpack}
\citep{Boggsetal1989} to perform orthogonal least squares regression, and
found a slope of 4.8$\pm$0.7. This fitting method uses the errors in $x$
and $y$ for each point, but is unable to take of account of the fact that
the errors are asymmetric. The mean of the upper and lower error is used
for the error in each axis.

This fitting method should be accurate as long as the points deviate from
the mean relation only on account of the statistical errors. In cases where 
there is a large intrinsic scatter about the relation, the statistical
errors do not provide an appropriate basis for the weighting of the data
points. An alternate approach is to weight all the points equally, ignoring 
the statistical errors as misleading. To check our result we also fitted
our \LT\ relation using the \textsc{slopes} package \citep{Isobeetal90} to
perform an OLS bisector fit and ignoring statistical errors. The best fit
slope for this technique was 5.05$\pm$0.44, very similar to the slope found 
when using the errors. 

From the temperature profiles for each galaxy, we are able to identify
which of our targets have strong temperature gradients which could affect
the measured mean temperature. Excluding these 20 galaxies weakens 
the relation to
$\sim$2.8$\sigma$ significance, and gives a best fit slope (fitted using
statistical errors) of 5.9$\pm$1.3. We
also fit the sample of galaxies with known temperature gradients, and found
a $\sim$2.7$\sigma$ correlation, with a slope of $\sim$3.7.

\begin{figure*}
\centerline{\epsfig{file=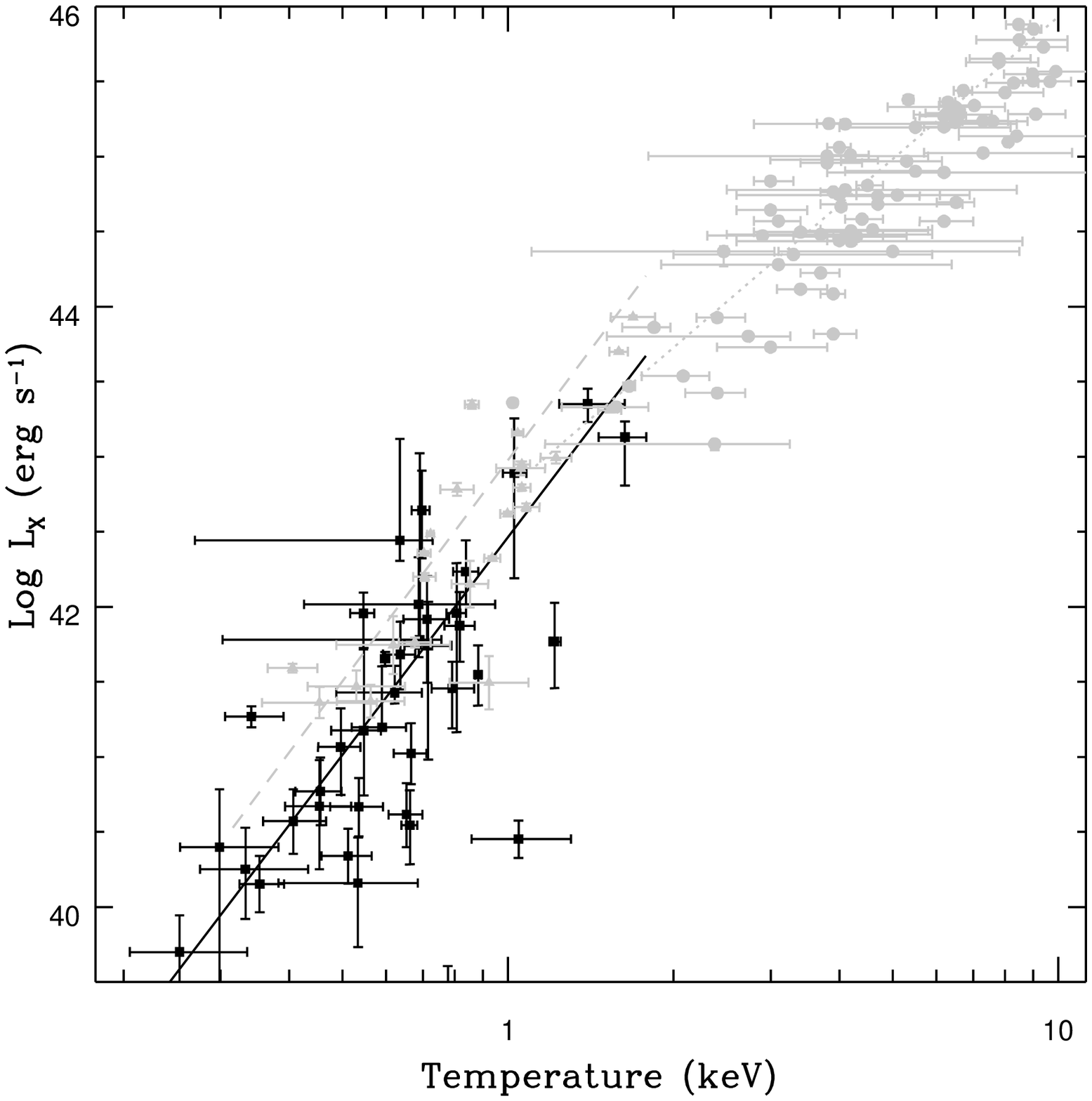,width=6in}}
\vspace{-5cm}
\caption{\label{fig:LTcomp} The \LT\ relation for our sample, with
  similar relations for groups and clusters. Our galaxy data points
  are marked (in black) by squares, and the relation by a solid
  line. Group and cluster data are shown in grey. Group data are taken
  from \protect\citet{Helsdonponman00}, and are marked by triangular
  points and a dashed line. Cluster data are drawn from
  \protect\citet{Davidetal93}, \protect\citet{MushotzkyScharf97} and
  \protect\citet{Fairleyetal00}, and are represented by circles and a
  dotted line. We find that the galaxy relation has a similar slope to
  that for groups, but is offset to fainter luminosities by a factor
  of $\sim$3}
\end{figure*}

The \LT\ relation has been used extensively in the study of groups and
clusters of galaxies. Figure~\ref{fig:LTcomp} shows our data points plotted 
alongside those for samples of groups \citep{Helsdonponman00} and clusters
\citep{Davidetal93,MushotzkyScharf97,Fairleyetal00}. Our data follow a
relation of similar slope to that of the groups (Helsdon \& Ponman find a
best fit slope of 4.9$\pm$0.8), but offset to a lower luminosity or higher
temperature. For a given temperature, our galaxies are a factor of $\sim$3
less luminous. Because of the scatter in both samples, there is some
overlap between the groups and galaxies, and some of our galaxy data points 
lie above the group best fit line. Conversely, the best fit relation for
clusters is significantly shallower than that for groups or galaxies,
though again there is a small region of overlap between the most luminous
galaxies and the faintest clusters.

\subsection{\LxLbtwo\ and \LbT\ relations}

\begin{figure*}
\centerline{\epsfig{file=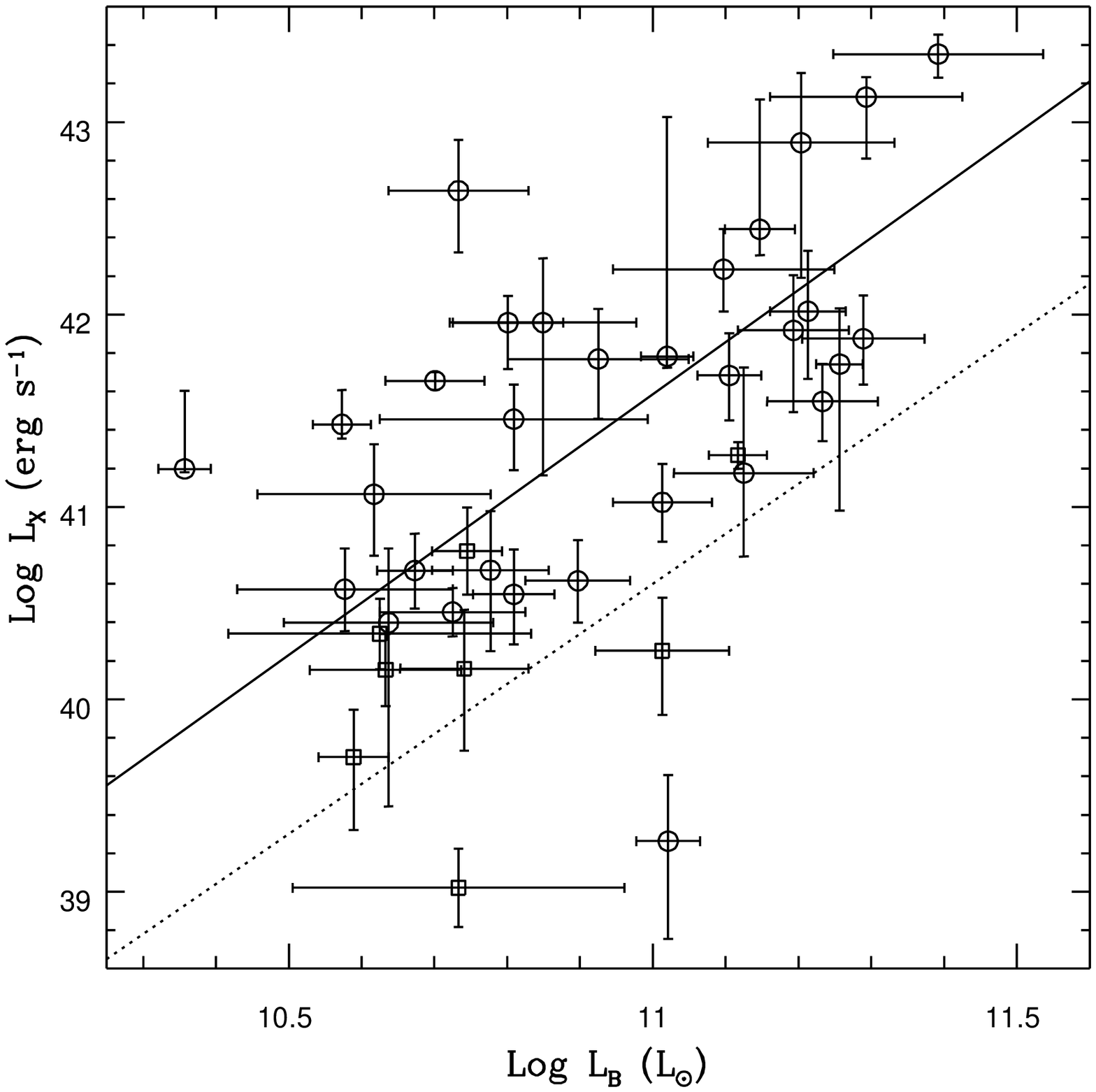,width=6in}}
\vspace{-5cm}
\caption{\label{fig:LXLB} \Lx\ plotted against \LB\ for our sample of
  galaxies. Although there are some outlying points, a clear trend is
  visible. Galaxies in X--ray bright groups are marked by
  circles, those in X--ray faint groups by squares. The solid
  line marks our best fit 
  relation, while the dotted line shows the best fit relation for X-ray
  bright groups \protect\citep{HelsdonPonman01}.}
\end{figure*}

One of the more common relations used in studies of early-type galaxies is
the \LxLbtwo\ relation. Numerous studies based on \rosat, \asca\ or
\einstein\ data have been published
\citep[\textit{e.g.}][]{Beuingetal99,Matsushita00b,Brownbreg98,Fabbianokimtrinchieri92},
and we have previously examined this relation in some detail in
\citet{OFP01cat}, to which we direct readers for a full discussion of the
relation and the effects of galaxy environment on it. Figure~\ref{fig:LXLB}
shows the \LxLbtwo\ relation for our galaxies. For the sample as a whole,
there is a 3.9$\sigma$ correlation, with a slope of $\sim$2.7. This is
quite a steep relation, comparable to that found for a sample of BGGs in
previous work \citep{OFP01cat}. We have also plotted the best fit relation
found for X-ray bright galaxy groups, from \citet{HelsdonPonman01}. The
slope of this relation (2.6$\pm$0.4) is very similar to that found for our
sample of galaxies. However, our relation is offset from that for groups,
with galaxies having X--ray luminosities a factor of $\sim$9 higher than
those of groups with equal optical luminosity, or conversely \LB\ values
$\sim$2.3 times lower than groups of similar \Lx. This result can be
compared to the \LT\ relation shown in Section~\ref{sec:LT}, in which
galaxies are offset to lower luminosities at a given temperature, compared
to groups.

\begin{figure*}
\centerline{\epsfig{file=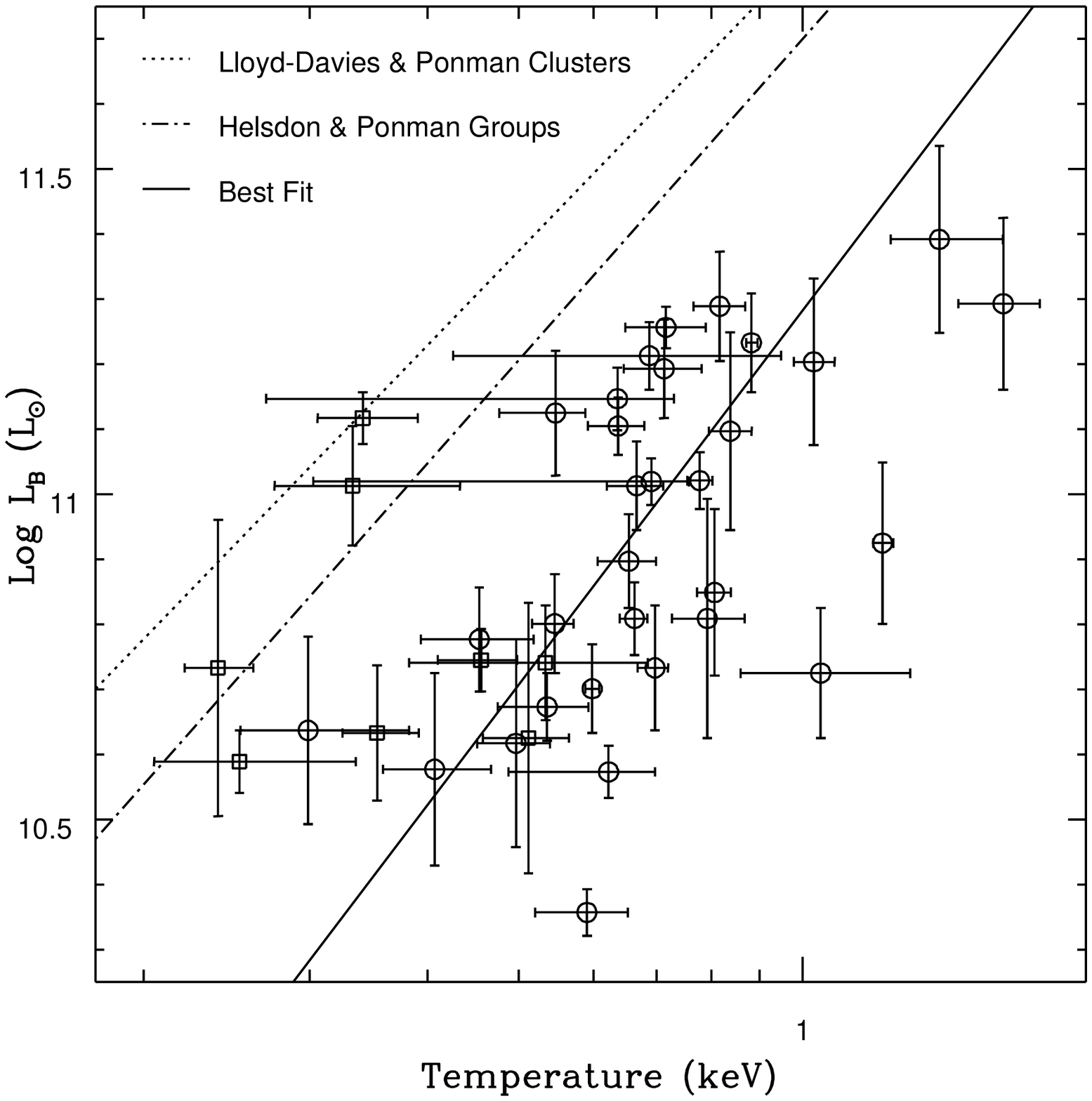,width=6in}}
\vspace{-5cm}
\caption{\label{fig:LbT} \LB\ plotted against temperature for our sample
  of galaxies. Circles mark galaxies which lie in X--ray bright groups
  or clusters, squares those in X--ray faint groups. The solid line is the
  best fit relation, while the 
  dotted and dash-dotted lines show the relations for clusters
  \protect\citep{Lloyd-DaviesPonman02} and X-ray bright groups
  \protect\citep{HelsdonPonman01} respectively.}
\end{figure*}

The \LbT\ relation for our galaxies is shown in
Figure~\ref{fig:LbT}. Once again, for this relation we find a fairly strong 
correlation ($\sim$4$\sigma$ significance). The slope of the relation is
comparable to that found for groups and clusters; 1.91$\pm$0.33 for our
galaxy sample, 1.64$\pm$0.23 for galaxy groups \citep{HelsdonPonman01}, and 
$\sim$1.5 for galaxy clusters \citep{Lloyd-DaviesPonman02}. However, where
the relations for groups and clusters are essentially the same
\citep{HelsdonPonman01}, our relation for galaxies is significantly offset
to higher temperatures (by a factor of $\sim$2, compared to groups) or
lower \LB\ (by a factor of $\sim$3). 

\subsection{The \sigT\ relation}
\begin{figure*}
\centerline{\epsfig{file=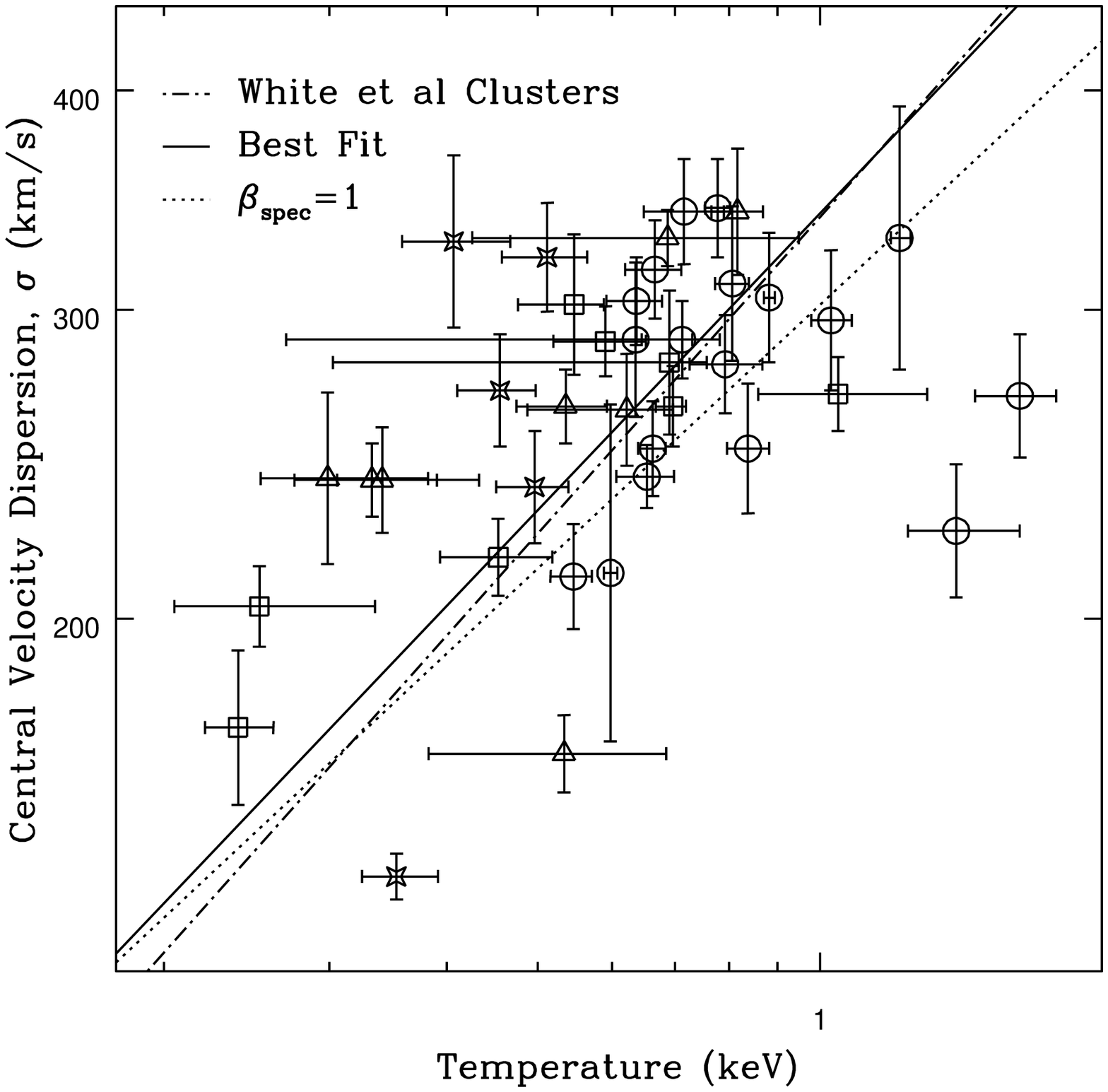,width=6in}}
\vspace{-5cm}
\caption{\label{fig:sigmaT} Velocity dispersion, $\sigma$, plotted against
  temperature for our sample of galaxies. Galaxies which show signs of
  central cooling are marked with open circles, those with central heating
  by stars, relatively isothermal galaxies by squares and those
  we have been unable to classify by triangles. 
  The best fit line to the sample as a
  whole is marked by a solid line, and the fits to the two subsamples are
  marked by dashed lines. $\beta_{spec}$=1 is marked by a dotted line,
  while the best fit relation for galaxy clusters
  \protect\citep[from][]{Whiteetal99} is marked by a dot-dashed line.}
\end{figure*}

For undisturbed objects with hydrostatic halos, both X-ray temperature
and velocity dispersion should be estimators of total system mass. It
should be noted that the $\sigma$ quoted for our galaxies is a stellar
velocity dispersion measured in the core of each object, whereas
similar relations for groups and clusters use the velocity dispersions
of the galaxies within those structures. Figure~\ref{fig:sigmaT} shows
the \sigT\ relation for our sample, again subdivided by temperature
structure. For the sample as a whole, we find a $\sim$3.1$\sigma$
correlation, with a slope of $\sigma \propto
$\Tx$^{0.56 \pm 0.09}$. However, there appears to be a large degree of
scatter about this line. Using the errors on the data points to
measure the expected statistical scatter, we can estimate the
intrinsic scatter of the data. We find that the data points are
$\sim$1.8 times more scattered than would be expected from the
statistical errors alone, hence the statistical scatter accounts for
56\% of the variance we see.

Also marked on the plot is the best fit relation for
clusters of galaxies, taken from \citet{Whiteetal99}. Our relation is
indistinguishable from this relation in both slope and intercept. The
relation for galaxy groups is somewhat steeper; \citet{Helsdonponman00a}
find a slope of $\sigma \propto$ \Tx$^{1.7 \pm 0.3}$ for their sample. Finally,
Figure~\ref{fig:sigmaT} shows a line representing $\beta_{spec}$=1, which
is expected for systems where there is equipartition of specific energy
between stars and gas. Our relation is consistent with this line, within
the errors.

Although there are many previous studies of samples of early--type
galaxies, very few have measured temperatures for their targets.
\citet{Daviswhite96}, one of the few exceptions, fit a \sigT\ relation to a
sample of 26 galaxies observed using the \rosat\ PSPC or \einstein\ IPC.
They find a slightly steeper relation than ours, with $\sigma \propto$
\Tx$^{0.69\pm0.1}$. Adding the errors in quadrature we see that their measured slope is comparable with ours, at the 1$\sigma$ confidence level. 
Their best fit line is however offset to higher
temperature. We believe this is probably caused by contamination by
surrounding emission or discrete sources, as the spectral fits use a
simple one component Raymond--Smith model \citep{Raymondsmith77}. Unless
the galaxies in their sample are completely dominated by halo emission, we
would expect such contamination to raise the measured temperature of single 
component models, and we found evidence of such biases when comparing one
and two component fits to galaxies in our own sample.

\subsection{\Bfit\ and Entropy}
\label{res:bfit}
Simple self--similar models of dark matter halos predict that X-ray
emission from gas in the halo will always take the same form. Assuming the
hot gas in the system is in hydrostatic equilibrium, the gas density can be
represented by a King profile \citep{Cavaliere76}. \Bfit\ values for groups
and clusters typically lie between 0.4 and 1. Studies of low mass clusters
and galaxy groups show that their surface brightness profiles become
shallower with decreasing \Tx\ \citep{Ponmanetal99}. This is usually taken
as an indication of additional physical processes affecting the gas, the
result of which is that the the gas halo appears more extended and diffuse.
Self--similar models also predict that gas entropy will vary linearly with
system temperature, as entropy is here defined as $S$=\Tx$/n_e^{2/3}$
(where $n_e$ is the electron density of the plasma), and mean gas density
will be constant for all systems which virialised at the same redshift.
For high mass clusters, this prediction matches the observed relation, but
in lower mass systems the behaviour alters, with the trend flattening so
that systems of temperature $\sim$1 keV seem to have entropy values
(measured at 0.1 R$_{virial}$) scattered around a mean value of $\sim$140
keV cm$^2$ \citep{Lloyd-Daviesetal00}.

The most common suggested processes which could affect the gas halo are
heating, either by galaxy winds \citep{Ponmanetal99} or AGN
\citep{Wuetal00}, or cooling of very low entropy gas
\citep{Muaonwongetal01}. \citet{VoitBryan01} have recently suggested that a 
combination of cooling and star formation is likely to be a highly
efficient way of producing the observed effects, as the heating will be
centred in regions containing the lowest entropy gas, giving the maximum
entropy increase. The existence of an entropy ``floor'',
suggests that the level of entropy increase may be similar
over a wide range of systems. In more massive systems, the entropy increase 
from shock heating is much larger than this amount, so it passes
unnoticed. Only in small systems does it become the dominant
contribution. Given this, and the fact that the suggested methods of
raising the entropy are likely to occur predominantly within galaxies, we
might expect that galaxy halos would show the effects of entropy increase
very clearly.

The entropy floor observed in low mass systems also has an effect on their
surface brightness profiles. Raising the entropy of the intra--cluster
medium (ICM) through heating will push gas out to larger radii if it occurs
after the system forms, or prevent it from collapsing as far as expected if
it occurs beforehand. The halo will therefore be more extended, with a
lower central density, presenting a flatter surface brightness profile. We
therefore expect \Bfit\ to decrease with decreasing temperature, and
this is observed across a wide range of systems 
\citep{Ponmanetal99,Lloyd-Daviesetal00}. Given
the likelihood that entropy increase has occurred in galaxy
halos, we may also expect to find a relation between \Bfit\ and
temperature.

\begin{figure*}
\parbox[t]{1.0\textwidth}{\vspace{-1em}\includegraphics[width=9cm]{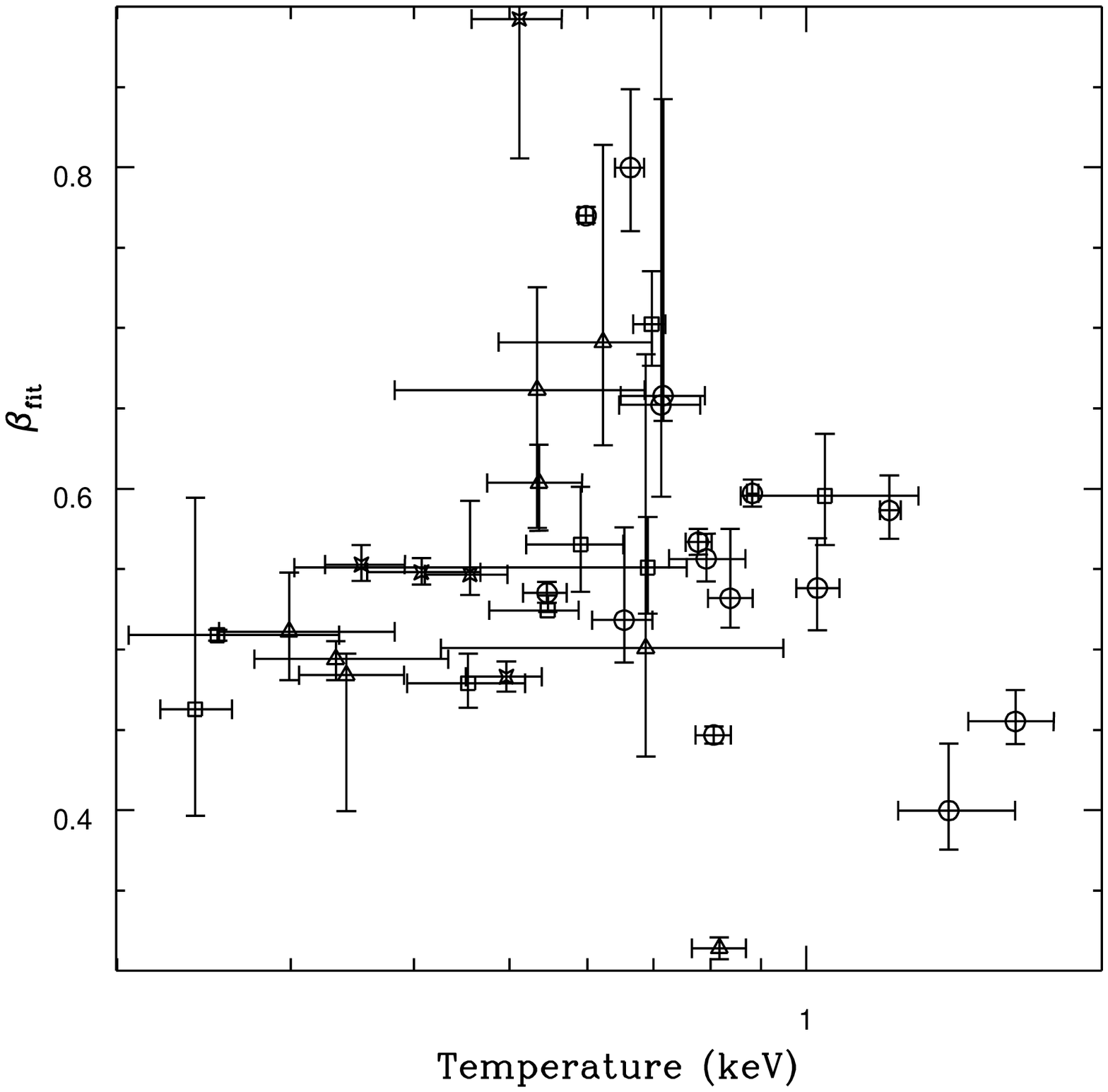}}
\parbox[t]{0.0\textwidth}{\vspace{-12cm}\includegraphics[width=9cm]{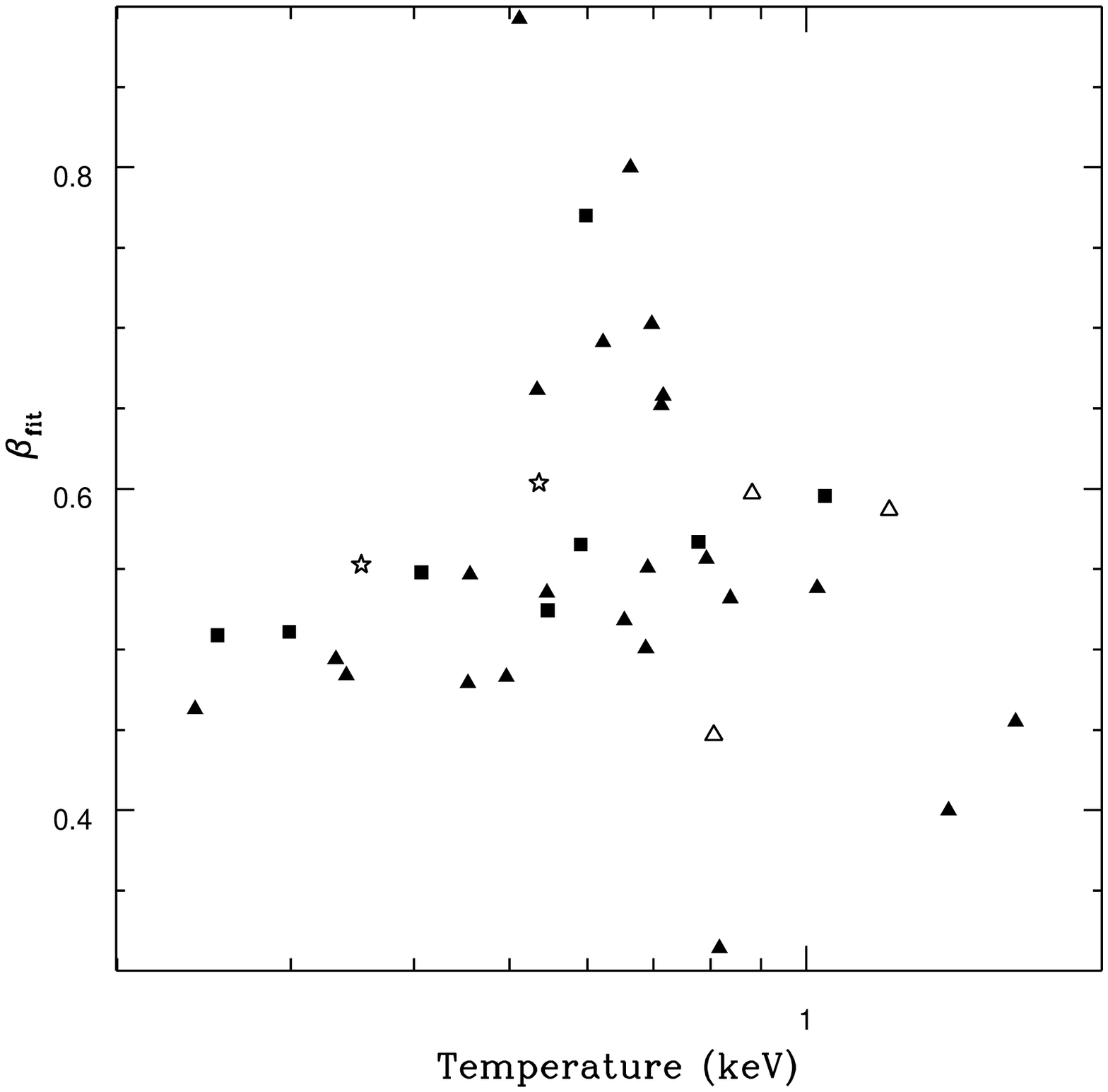}}
\vspace{-3cm}
\caption{\label{fig:BetaT_CF} Plots of \Bfit\ against temperature for 
  our sample of galaxies. On the left, galaxies are grouped by their
  temperature properties, with those with cool cores marked by open
  circles, those with hot cores by stars, those with isothermal profiles by
  squares, and those we have been unable to classify by triangles. On the
  right we segregate galaxies depending on their environment, with filled
  triangles representing brightest group galaxies, open triangles brightest
  cluster galaxies, stars AGN, and squares normal early-type galaxies.
  Error bars are omitted for clarity.}
\end{figure*}

Figure~\ref{fig:BetaT_CF} shows the slope parameter \Bfit\ plotted against
temperature for our sample. There is no obvious trend in the points, and we
find no statistically significant correlation.  The scatter on the points
is quite large, particularly at intermediate temperatures. There is no
clear segregation of galaxies by temperature structure, and all classes
show comparable amounts of scatter. There is some suggestion that AGN, BGGs
and BCGs are more scattered than the more normal ellipticals, particularly
if the normal galaxy with the highest value of \Bfit\ is excluded. This
galaxy is NGC~1404, which is commonly considered to have suffered
ram-pressure stripping of its halo. The sharp cut off in its surface
brightness profile could conceivably have produced an unusually steep fit.
However, the small size of the subsample makes any detailed comparison
unreliable. The subsample seems to be centred on $\beta_{fit}\simeq$0.55,
which is similar to the mean value of our sample as a whole.

Assuming that our galaxies are isothermal, we can extrapolate from our
2-dimensional surface brightness models to 3-dimensional density models of
our galaxies. Equation~\ref{eqn:kingSB} describes the 2D models and
we can describe the 3D models as shown
in Equation~\ref{eqn:king3d}.
\begin{equation}
\rho_{(r)} = \rho_{(0)}[1+(r/r_{core})^2]^{-\frac{3\beta}{2}}
\label{eqn:king3d}
\end{equation}
The density normalisation, $\rho_{(0)}$, can then be determined from the
surface brightness normalisation, assuming the temperature and metallicity
determined from spectral fitting. From these 3D models,
it is possible to derive gas properties such as density, entropy, cooling
time, etc, as a function of radius. In order to be able to compare the
resulting profiles fairly, we need to view them on a common scale. This can
be done by scaling the profiles by the virial radius of the system, or more
usually to fixed overdensity radius, such as R$_{200}$. The overdensity is
calculated relative to the critical density of the universe at the redshift
of formation. This is unknown for most systems, and in the case of clusters
and groups is generally taken to be the redshift of observation. For
galaxies, such an assumption is very unlikely to be accurate, even taking
into account the fact that early-type galaxies will have potentials
determined by the density of the universe at the redshift of last major
merger rather than that at which the majority of their stellar population
formed. In order to calculate R$_{200}$, we assume a mean redshift of
formation for ellipticals of $z_{form}$=2
\citep{Kauffmannetal96,Vandokkumfranx01}.  Using this value, we can
calculate R$_{200}$ as described in \citet{Baloghetal99} and
\citet{Babuletal01}, taking variation of overdensity with redshift from
\citet{Ekeetal96}.

\begin{figure*}
\centerline{\epsfig{file=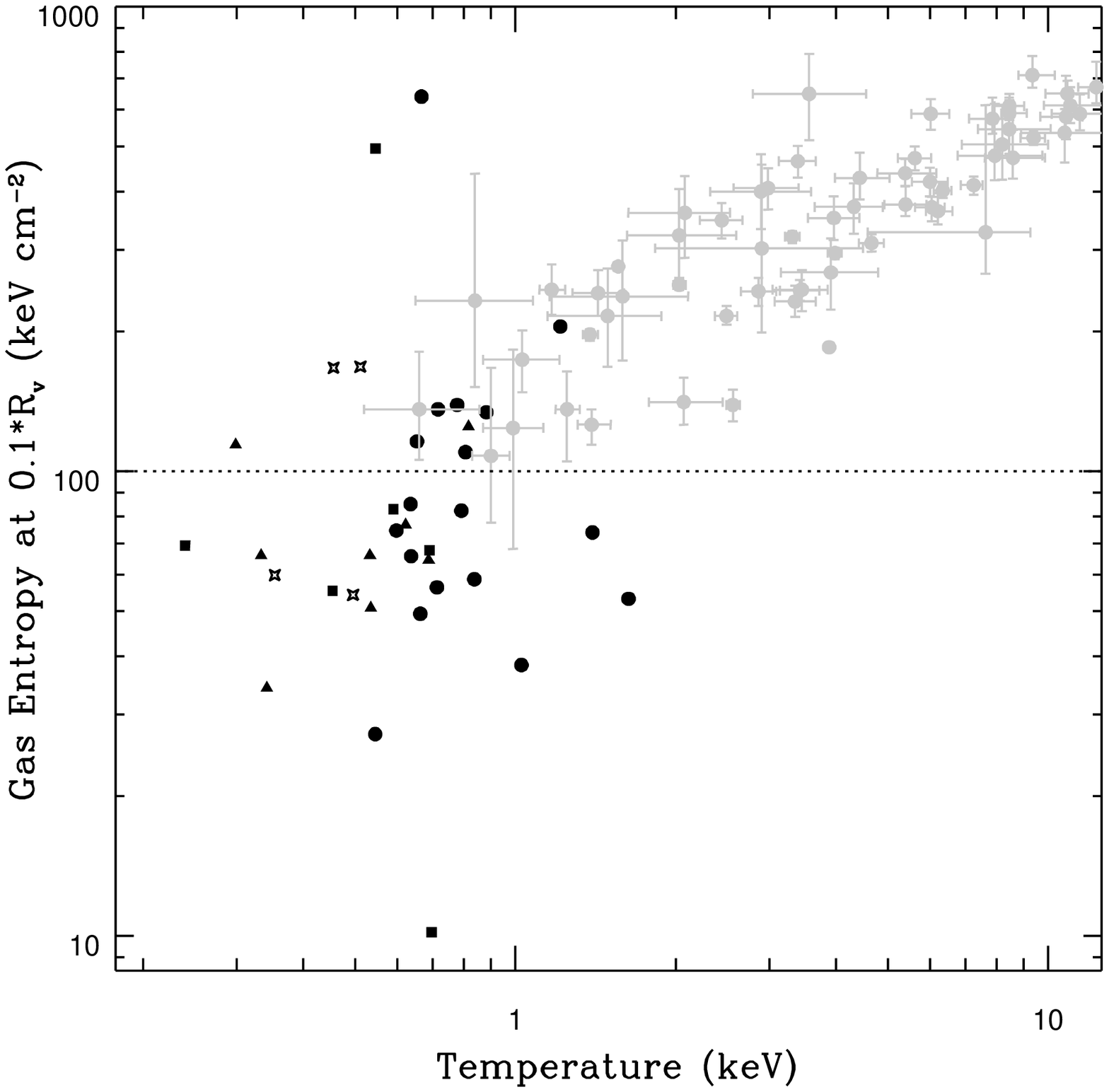,width=6in}}
\vspace{-5cm}
\caption{\label{fig:entsig} Mean gas entropy measured at one tenth of the
  virial radius plotted against mean halo temperature. Galaxies with
  cool cores are marked as circles, galaxies with relatively
  isothermal halos by squares, those with hot cores by
  stars and those we were unable to classify by triangles. Grey
  points with error bars represent groups and clusters taken from the
  Birmingham-CfA cluster scaling project
  \protect\citep[see][]{Sandersonetal02}. The approximate level of the
  entropy floor for groups and clusters is marked by a dotted line.}
\end{figure*}

Figure~\ref{fig:entsig} shows entropy, calculated at one tenth of
R$_{200}$. Once again, galaxies are marked to show their different
temperature structures, but while the data shows a large amount of scatter,
there is no significant correlation with temperature, or segregation by
temperature structure. The galaxy data points do mainly fall at the low
temperature end of the \ST\ trend for groups and clusters, but show no
evidence of any trend themselves.

\section{Discussion}
\label{sec:discuss}

The results presented in the previous section do not lend themselves
to a simple explanation. The halos of massive early-type galaxies seem
to have some obvious similarities to those of galaxy groups and
clusters, but also some intriguing differences. Clearly a number of
the relations which hold for groups and clusters also apply to
early--type galaxies, the most important examples being the \sigT\ and
\LT\ relations. The \sigT\ relation is particularly striking, in that
it agrees closely with the relation found for galaxy clusters. The
\LT\ relation has a similar slope to that of galaxy groups, but is
offset to lower luminosities. The \LbT\ and \LxLbtwo\ relations also
show offsets from the cluster and group relations, but to higher
X--ray luminosity and temperature.  Another important result is the roughly
constant value of \Bfit\ across a wide range of temperatures. Although
there is some suggestion of influence by surrounding groups and
clusters on this parameter, the environment seems to increase scatter
rather than producing a trend. Together with the results of our
entropy calculations for the sample, the behaviour of \Bfit\ suggests
that if preheating effects are important in galaxy halos, they produce
results quite different to those seen in groups.

\subsection{Comparison with Groups and Clusters}
One of the most important results to arise from studies of galaxy groups
and clusters is the demonstration that simple self-similar models do not
describe low mass systems well. There is good evidence that galaxy groups
differ from higher mass clusters in a number of important ways. The most
widely accepted explanation for these differences is that the hot gas in
the halos of these systems is affected not only by the processes involved
in formation of the system as a whole, but also by processes such as star
formation, AGN heating, and gas cooling. A clear sign of this is seen in
the behaviour of gas entropy in systems of different mass. In high mass
clusters, entropy can be fairly accurately predicted from simple models in
which the gas is heated (and has its entropy profile set) during formation
of the system. As the cluster builds up, gas flows into the potential well
and is shock heated to a degree dependent on the depth of the potential.
This dependence of halo temperature on system mass can be seen in the
\sigT\ relation for clusters, in which \Tx\ is strongly correlated with
$\sigma$, with \Tx\ \simpropto\ $\sigma^2$, as expected from arguments
based on the virial theorem.

However, as system mass (and temperature) drop, the gas entropy begins to
depart from the predicted relation, with low mass galaxy groups having
higher than expected gas entropies. \citet{Lloyd-Daviesetal00} have shown
that at temperatures below $\sim$2 keV, gas entropy appears to remain
roughly constant, scattering about a mean value of 140 keV cm$^2$. An
effect is also seen at around this temperature in the surface brightness
profiles of galaxy groups and clusters \citep{Ponmanetal99}. High
temperature systems (\Tx\ $>$ 4 keV) have profiles which scale
self-similarly, but as \Tx\ is lowered, the surface brightness profiles
observed are found to be shallower, with lower central densities. This has
an obvious effect on the measured X--ray luminosities of these systems,
making them fainter than predicted.

Both of these trends are explained through the effects of
non-gravitational processes on the gas. To take a simple model, we can
imagine a system at the time of formation. As stated above, we expect
the gas falling into the potential of the system to be heated by
gravitational processes, but we might also expect heating from other
sources. For example, as galaxies form in the system we could expect
star formation, galaxy winds, and AGN to affect the gas in the system
halo. In a large system, the contribution from these sources will be
small compared to that of gravitational heating.  However, in smaller
systems it will be more important, and will eventually dominate the
energy of the gas halo. This additional heating will have a number of
effects, such as raising the gas temperature, causing the halo to
expand by moving gas to higher radii, raising the entropy of the gas,
and so on. In practice, it is possible to achieve these effects
through a number of processes, or a combination of several. One
promising model is that of \citet{VoitBryan01}, in which low entropy
gas cools rapidly, providing material for star formation. This star
formation then not only removes low entropy gas from the system
(raising the mean entropy), but provides heating which is focused in
the areas in which low entropy material dominates. From our point of
view however, the method is not important; all the suggested processes
would occur in galaxies, and might be expected to occur preferentially
in large galaxies at the bottom of a group or cluster potential
well. For simplicity, we will refer to this as the preheating model.

\subsubsection{The \sigT\ relation}
\label{sec:disc_sigt}
One of the clearest similarities between the relations for our galaxies and
those of larger structures is the correspondence of the \sigT\ relation to
that of galaxy clusters. As discussed above, the correlation in clusters is
expected, and shows that both \Tx\ and $\sigma$ are good measures of the
potential. In galaxy groups, the slope of the relation is observed to be
somewhat steeper, $\sim$1.7$\pm$0.3 \citep{Helsdonponman00a}. One
explanation for this steeper slope is that \Tx\ is raised in low mass
systems by the heating and removal of cool gas described above. An
alternative is the suggestion that the velocity dispersions measured in
groups could be biased. There are a number of groups which, despite
apparently possessing extended X-ray halos, have exceptionally low velocity
dispersions \citep{Helsdonponman00}. If these $\sigma$ values were
accurate, the potential of the group would not only be too shallow to produce
the halo luminosity observed, but would be too shallow for the group to
have collapsed within the age of the universe. Several reasons for
underestimation of $\sigma$ can be postulated. It is possible that these
low mass groups form a central core of bright galaxies, with fainter
members at higher radii. The brightest galaxies are most likely to be
recognised as group members and have measured redshifts, so they will
dominate any calculation of $\sigma$. Tidal interactions between group
members could reduce velocity dispersion by transferring orbital energy
from the galaxies to their stars.  It has also been suggested that low mass
groups may be biased because the groups often have prolate structures. For
groups whose major axis lies near the plane of the sky, this particular
structure could lead to an underestimation of $\sigma$
\citep{Tovmassianetal02}, biasing samples including such low temperature
systems. If $\sigma$ is biased, correcting the measurements for the lowest
temperature groups would probably shift the best fit relation into
agreement with that seen in clusters. It is worth noting that the groups
with the lowest velocity dispersions listed in \citet{Helsdonponman00} are
also amongst the poorest; all groups in their sample with $\sigma <$ 150
\kmps\ have only 3-5 member galaxies. Calculating velocity dispersion from
such small numbers can introduce a bias, producing values which are
underestimated by up to 15\% \citep{Helsdon02}.

The fact that our results agree well with the cluster relation suggests
that again, we are looking at systems in which both \Tx\ and $\sigma$ are
good measures of the potential. From a preheating point of view this is
surprising. If we expect preheating processes to occur in large galaxies,
then we might expect the \Tx\ of their halos to be raised, in much the same 
way as we see in galaxy groups, producing a steeper relation. We also need
an explanation of why the gas in these galaxies has a temperature which is
related to the depth of their potential well. As well as additional heating
from star formation and AGN activity, we expect large quantities of gas to
be lost from the stars in the galaxy. As this gas is produced within the
galaxy, we cannot expect shock heating during infall, so we might initially 
believe its temperature to be entirely determined by supernova heating.

\citet{Helsdonetal01} consider a related question, that of  
the relative importance of different energy sources which contribute to the
total X--ray luminosity of early--type galaxies. Their Figure~8 is a plot of
\LxLb\ against \LB, showing the contributions to \Lx\ from discrete sources,
SNIa and gravity. From our point of view the contribution from
discrete sources is irrelevant, as we are interested in energy input to the 
gas in the galaxy. The gravitational contribution is a combination of two
processes, firstly a contribution from the velocity of the stars in the
potential (gas lost from these stars will have an added kinetic energy
component from their velocity, which will be thermalized in the surrounding 
ISM), and secondly a contribution from work done on the gas as gravity
causes it to contract and cool. The important result with regards to our
situation is that while the SNIa energy input scales with \LB, the
gravitational input scales with $\sigma^2$. This means that for low mass
systems, the dominant contribution to \Lx\ is from supernova heating, but
above \LB\ $\sim$ 10$^{10}$~\Lsol\ gravitational work begins to
dominate. If we consider these two factors as energy inputs to the gas
rather than as contributors to \Lx, we can see that the gas temperature
is likely to be determined by the supernova rate in low mass systems, and by the
depth of the potential in high mass systems. The point at which the two are 
equal will depend on the details of the model, but if we follow the
assumptions made by \citet{Helsdonetal01}, all our galaxies lie in the high 
mass, gravitationally dominated region of the plot. \textbf{We could therefore
expect gas temperature to depend on the depth of the potential, whether the
gas has an internal or external origin.}

It seems likely, from this result, that we can draw similar conclusions
from the \sigT\ relation in galaxies as we would in clusters. \Tx\ and
$\sigma$ are both probably good measures of mass. Preheating, by whatever
method, does not appear to have the effect on galaxy halos that it has on
those of groups. Like clusters, the galaxy relation is consistent with the
systems having $\beta_{spec}$ = 1, suggesting that there is equipartition
of specific energy between stars and gas. As we will discuss later, this
has important implications, in that it suggests that the optical and X--ray
density profiles should be similar. One further consideration is the
intrinsic scatter in the points. Our data has $\sim$1.8 times as much
scatter as we would expect from statistical errors alone, giving us a
non-statistical unceratinty in T of $\pm$0.04 for any given value of
$\sigma$. For comparison, the galaxy groups studied by
\citet{Helsdonponman00a} are less scattered, having only $\sim$1.4 times as
much as would be expected from the errors, sothe non-statistical
uncertainty in T for these systems is $\pm$0.86. The degree of scatter in
the galaxy data could have a number of causes. Gas temperature could be
affected by many processes, related to the galaxy or the surrounding
environment.  Velocity dispersion could also be affected by processes
associated with the formation or merger history of the galaxy. However, it
would appear that galaxy groups have a larger scatter in properties,
suggesting that the galaxies are less affected by external influences.

\subsubsection{The \LT\ relation and \Bfit}
\label{sec:disc_LT}

Accepting $\sigma$ and \Tx\ as indicators of the mass of the system, we
next consider the \LT\ relation. Here, we find that the slope of the
relation is steeper than that of clusters, as steep as that of groups. The
relation is also offset, so that at a given temperature, galaxies have a
lower luminosity than groups. The steep slope is usually explained as a
product of preheating - the additional heating of the gas raises the
temperature slightly and moves gas to higher radii, lowering the central
density and therefore \Lx. If we assume that the steep slope seen in
galaxies is caused by the same processes which cause it in groups, then
this relation is a strong piece of evidence for the effects of preheating
in galaxies.

However, there are other ways in which we might produce such a steep slope.
\citet{Helsdonetal01} found a strong correlation between the X--ray
properties of group dominant galaxies and those of the groups in which they
are found. Given the signs of central cooling in many of these groups, they
suggested that what had been initially identified as the halos of the
dominant galaxies were in fact group scale cooling flows, centred on the
dominant galaxy because it lies at the bottom of the group potential. They
also found that the \Lx\ of this central galaxy halo/cooling flow was
$\sim$25\% of the \Lx\ of the group. As 29 of our 39 galaxies are dominant
galaxies in groups, clusters or cluster subclumps, we must consider the
idea that our relations could be dominated by cooling flows. In that case,
we would expect the \LT\ relation to have a similar slope to that of groups,
but to be offset to lower \Lx\ values by a factor of 4, and also to lower
temperatures. Provided the temperature drop is not too large, this could
reproduce the relation we see very well. On the other hand, we do not see
any segregation in the data between those galaxies which show signs of
central cooling and those which do not, nor do we see any difference
between galaxies at the centres of groups and those in other environments.
This argues against cooling flows as the driver of the relation. We will
discuss the evidence for and against cooling flows as the dominant factor
in Section~\ref{sec:disc_CF}.

The lack of a relation between \Bfit\ and \Tx\ argues against both cooling
flows and preheating as the source of the \LT\ relation. In galaxy groups,
preheating causes gas to move out to high radii, reducing \Bfit\ as the
surface brightness profile becomes flatter. As preheating is more effective
in smaller mass systems, groups and low mass clusters show a correlation
between \Bfit\ and \Tx\, with cooler systems having flatter profiles. In
galaxies, despite a very large scatter, we see no trend with temperature.
Our galaxies lie around a mean value of \Bfit\ = 0.55, which means that
even if the group and cluster \BfitT relation levelled off at low
temperature, our sample would not be consistent with it. \textbf{This
  strongly suggests that preheating is not the cause of the steep slope of
  the \LT\ relation.}

\begin{figure*}
\centerline{\epsfig{file=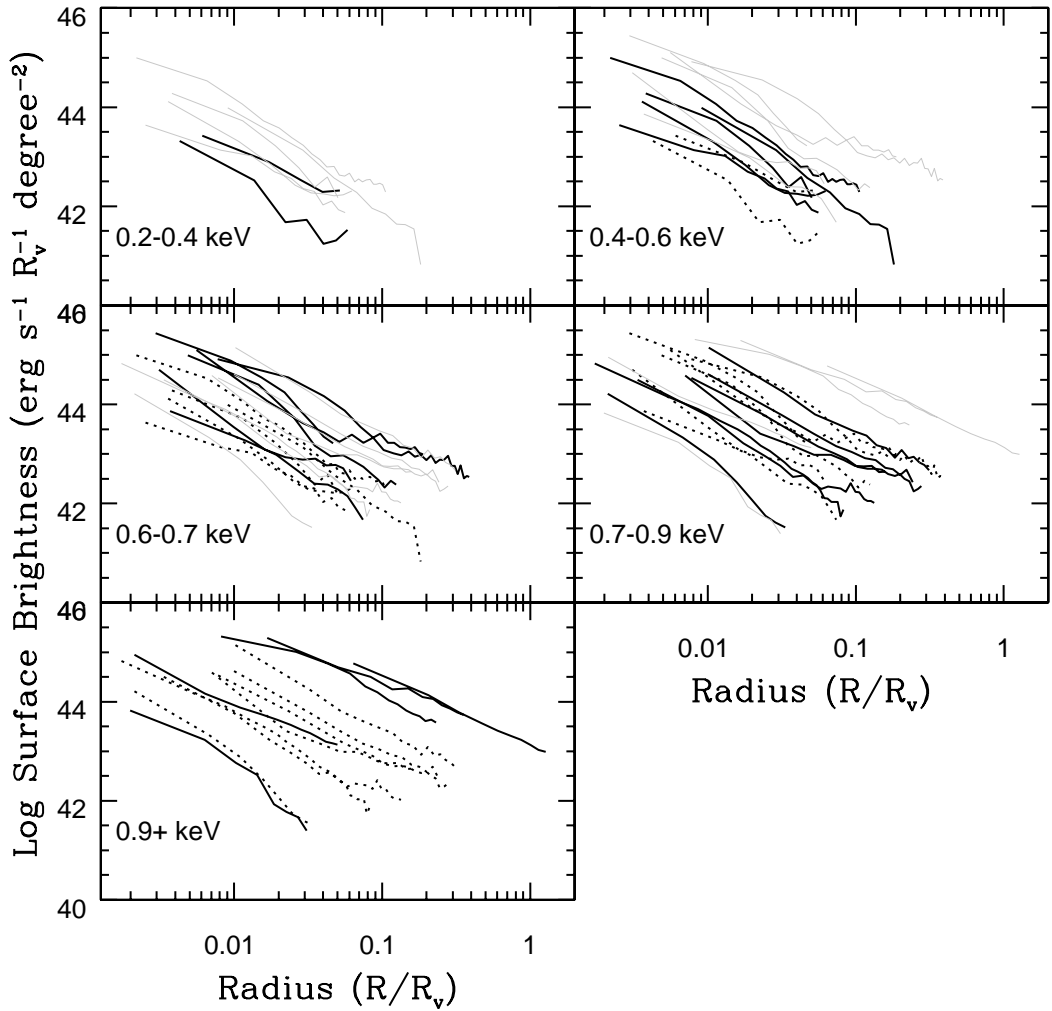,width=6in}}
\vspace{-5cm}
\caption{\label{fig:SBprofs} Scaled X-ray surface brightness profiles for
  our galaxies, arranged to show departures from self-similarity. The
  profiles have been scaled to the virial radii of the galaxies
  (assuming a redshift of formation $z_{form}$=2), and by \Tx$^{1/2}$
  to account for self-similar scaling with mass
  \protect\citep{Ponmanetal99}. We have also converted from counts to
  intensity, using the best fit MEKAL component of each galaxy to
  allow for the temperature and metallicity dependence of the plasma
  emissivity. As all our galaxies are at low redshift ($z <$ 0.035) we
  ignore cosmological effects as negligible. In each plot bold, dark lines
  show the galaxies in the energy band in question, narrow, grey lines show
  those in the band above, and dotted lines those in the band below. 
  Although there is some scatter, it
  is clear that all have similar slopes and that cooler systems have
  lower normalisations.}
\end{figure*}

If a change in the slope of the surface brightness profile is not
responsible for the drop in \Lx\ needed to produce the \LT\ relation, then
there must be a drop in normalisation. This is demonstrated in
Figure~\ref{fig:SBprofs}, which shows the surface brightness profiles of
our sample, scaled so that if they were behaving self-similarly, they would 
coincide. Details of this scaling are given in the figure caption. 
A comparison between Figure~\ref{fig:SBprofs} and Figure 1 of
\citet{Ponmanetal99} demonstrates the difference between the behaviour of
our galaxies and that of groups and clusters. The high temperature clusters
in Figure 1 of \citet{Ponmanetal99} behave self-similarly, having roughly
equal surface brightness slopes and scaled normalisations. The lower
temperature groups move away from self-similarity, as their slopes flatten
and central surface brightness (and density) is lowered. In the galaxies,
although there is a large amount of scatter, the slope remains roughly
constant across the range of temperatures, but the normalisation of the
profiles seems to drop with decreasing temperatures. \textbf{This suggests
  that whereas in galaxy groups the steepening of the \LT\ relation is
  caused by energy injection and movement of gas to higher radii, in
  galaxies is caused by changes in normalisation and therefore in the
  overall gas fraction of the systems.} The processes responsible for this
change are not clear, but this result demonstrates how differently galaxies
behave compared to groups and clusters.

It is interesting to note that the mean slope of the surface brightness
profiles (\Bfit~=~0.55) is similar to the mean slope of the optical
surface brightness profile. Elliptical galaxies are well described in the
optical by a de Vaucouleurs profile, which is similar (outside the central
regions) to a King model with \Bfit\ = 0.5. As mentioned in
Section~\ref{sec:disc_sigt}, the \sigT\ relation is consistent with
$\beta_{spec}$ = 1, which leads us to expect a similarity between the
optical and X-ray density profiles. The similarity in surface brightness
profiles is therefore further confirmation of the \sigT\ result. A further
interesting note is that galaxy wind models, in which the majority of the
gas in the halo is produced by stellar mass loss, predict X--ray surface
brightness profiles similar to those in the optical
\citep{Pellegriniciotti98}. The exact value of \Bfit\ would depend on the
wind state of the system, with only relatively hydrostatic halos having
\Bfit\ = 0.5. Galaxies dominated by supersonic outflows, or those in which
inflows are important, would be expected to have steeper profiles owing to
their high central gas densities. Galaxies dominated by subsonic outflows
would be expected to have slightly flatter profiles (\Bfit\ $<$ 0.5), as
the outflowing gas has a higher density at large radii than in a supersonic
flow, and less of a central peak. If such models are applicable, it seems
likely from the high mass of our galaxies that they would be in the inflow
stage, so we would expect \Bfit\ $>$ 0.5, in agreement with our
observations.

\subsubsection{Entropy}
The results of our entropy measurements are in broad agreement with those
from the surface brightness profiles. In galaxy clusters and groups 
we see evidence of similarity breaking and preheating, leading to a trend
in entropy which levels off at the entropy floor. The galaxy data points do 
not show any sign of a trend, and although they may appear to be consistent 
with the general trend in higher mass systems, the scatter in the data is
very large. Our method of calculating entropy has two important sources of
scatter associated with it. Firstly, we must assume that our galaxies are
isothermal, despite the fact that we know that many of them have
temperature gradients. Secondly, we measure the entropy at one tenth of the 
virial radius, and when calculating the virial radius we must assume a
redshift of formation (or last major merger). Although the assumed value of 
$z_{form}$ = 2 is probably a reasonable mean, there will clearly be
variation between individual galaxies. Our virial radii will therefore be
inaccurate to some degree, meaning that we are actually measuring entropy at 
a range of scaled radii.

Despite these difficulties, we can draw some important results from the
entropy values. A large proportion of the data points lie below 100 keV
cm$^2$, meaning that they are below the entropy floor observed in galaxy
groups.  One possible reason for these low values is that the galaxies form
earlier than the surrounding groups. The density of systems virialising at
a given epoch is related to the critical density of the universe at that
time, and so the earlier an object virialises, the denser we expect it to
be. As entropy is inversely proportional to $n_e^{2/3}$, an equal amount of
energy injected into a denser system will produce a smaller increase in
entropy. Galaxies, forming at $z$ $\sim$ 2, might therefore be expected to
have higher densities and a lower entropy floor than groups and clusters
which have formed more recently. However, we might also expect that as the
site of the processes responsible for entropy increase, the amount of
energy available to affect entropy in galaxies would be larger than in
groups. It is also worth considering that if galaxies had formed with halos 
of very low entropy gas, it would have cooled on timescales considerably
shorter than the Hubble time, probably leading to star formation, heating
and a rise in observed entropy \citep{VoitBryan01}.

An alternative viewpoint is that the majority of the gas in these
systems is being produced by stellar mass loss rather than infall, so
we should not expect entropy to behave as it does in groups. In this
case, galaxy wind models should give a reasonable approximation of the
behaviour of the halo, in the absence of significant environmental
influences. Several published simulations of galaxy halo development
give gas density and temperature profiles
\citep{Ciottietal91, Pellegriniciotti98, BrigMath99} for their model
galaxies. In models of outflowing winds, both gas density and
temperature fall with radius, but density falls more rapidly, so we
would expect an entropy profile which increases with radius. In models
which have developed gas inflow (cooling flows), temperature may rise
with radius throughout most of the model, so again we expect entropy
to rise with radius. This matches what we observe in our measured
entropy profiles. The entropy predicted at any given radius depends on
the details of the model, \eg\ system mass, age, supernova rate, mass
injection rate, \etc Entropies ranging from 20-300 keV cm$^2$ might be
expected for galaxies such as those in our sample. The majority of our
galaxies do have entropies within these limits, and considering the
expected scatter, the agreement between models and measurements is
fairly good.

\subsection{Cooling Flows}
\label{sec:disc_CF}
A number of our galaxies have temperature profiles indicative of central
cooling. This is not surprising, as they are fairly massive objects, with
large halos, and reside within larger structures which are themselves
probably capable of producing cooling flows. Among the relations we have
examined, several show behaviour which could be explained easily as the
product of group cooling flows. The best example of this is the \LT\ 
relation, where the slope is identical (within errors) with that found for
groups, but offset to lower X-ray luminosities. As discussed in
Section~\ref{sec:disc_LT}, this offset (a factor of 3 in \Lx) is very
similar to that we would predict (a factor of 4), based on studies of X-ray
bright groups \citep{Helsdonetal01}. We would expect the measured \Tx\ of
the cooling flow region to be lower than the group temperature and this
could explain the difference between predicted and measured offset. We
might also be able to explain the offsets observed in the \LxLbtwo\ and
\LbT\ relation using this model. In both cases we are comparing a parameter
which is determined by the galaxy and only weakly influenced by the group
(\LB) to a parameter which is determined by the cooling flow and hence by
the properties of the group (\Lx\ or \Tx). In such a situation, we would
have to expect \Lx\ and \Tx\ to appear unusually high for the central
galaxy, as they would actually be related to the much larger system of the
surrounding group or cluster. Comparing with optical luminosity, we would
only expect to find a relation without an offset if we compared them to a
value of \LB\ calculated for all the galaxies in the group, rather than
just the central dominant elliptical. We would also expect the galaxies in
X--ray faint groups to behave differently, as their halos could not be the
product of group scale cooling flows. Unfortunately, the small numbers of
such objects in our sample means we cannot test whether they follow a
different trend to those in X--ray bright groups, but they do indeed appear
to fall at lower \Lx\ and \Tx, for a given \LB.

However, there are two powerful arguments against group scale cooling flows
as the dominating influence in our sample. Firstly, we find that a
significant subset (one third) of our sample do not show signs of central
cooling. They are instead approximately isothermal, or have a central rise
in temperature. This does not necessarily mean that they have not been at
the centre of a group cooling flow in the past, but such a flow would have
to have been disrupted, and so would be unlikely to produce the emission we
assume to be a halo. Secondly, and most importantly, we see no sign of a
difference between those galaxies with or without signs of central cooling.
The only segregation we see in any of the relations is a tendency for
galaxies which seem to harbour cooling flows to be higher mass systems, and
so to have higher values of $\sigma$, \Tx, \Lx, \etc We see no other
difference between galaxies with differing temperature profiles, and we
specifically do not see the galaxies with apparent cooling flows driving
the \LT\ relation. This strongly suggests that while a number of our
galaxies do harbour cooling flows, some of them quite large, their halos
are not simply cooling flows formed by surrounding groups and clusters,
unless the properties of the gas halo are independent of whether it arises
from stellar winds or group inflow.

\subsection{Stellar Mass Loss and Galaxy Winds}
Large scale cooling flows, in which most of the gas in the galaxy halo has
an external origin, do not appear to provide a good model for our galaxy
sample. The alternative is a model in which the majority of the gas is
produced (and heated) internally, via stellar mass loss. We have already
discussed in Section~\ref{sec:disc_sigt}, the relative contributions to
the energy of gas produced in a galaxy from supernovae and gravitational
processes. Our sample is made up of galaxies with high stellar masses,
leading us to believe that the gravitational potential should be the
dominant source of energy. This provides an important link between the
X--ray properties of the galaxies and their mass, which could explain the
correlations we see between optical and X--ray properties in our sample.
If gravity is the dominant source of energy in the halos of our galaxies,
we would expect them to behave like clusters on many of the relations we
have examined. The best example of this is the \sigT\ relation, where the
galaxies and clusters have best fit lines which are indistinguishable.

However, there are a number of ways in which the relations do not behave
like those of galaxy clusters, and we need to find explanations for these
differences. The most interesting, and perhaps the most important, is the
\LT\ relation. This behaves like that of galaxy groups, rather than galaxy
clusters. However, instead of a decrease in central gas density in low mass
systems, leading to a trend in \Bfit\ with temperature, we see a constant
value of \Bfit\ and a decrease in the normalisation of the surface
brightness profiles at low masses. This suggests that the lower mass
members of our sample lie progressively further below the cluster \LT\ 
relation because they have a lower gas fraction than their higher mass
counterparts. To explain the \LT\ relation we observe, we need an
explanation of this change in gas fraction with temperature.

A number of possible reasons for this trend in gas fraction (\Fgas) could be
suggested. These include:

\begin{enumerate} 
\renewcommand{\theenumi}{(\arabic{enumi})}
\item The change in \Fgas\ is a natural consequence of halo
  production by stellar mass loss. Any steady state solution would produce
  the trend we see.
\item The change in \Fgas\ is the product of the surrounding
  environment. Higher mass systems tend to be in higher mass groups and
  clusters, which have a denser IGM. This prevents the escape of gas, or
  compresses the halo, or allows accretion of gas, causing the halo to
  reach higher densities and gas fractions.
\item The lower gas fraction in cooler galaxies is an evolutionary effect,
  related to the age (or time since last major merger) of the galaxy. The
  mass density of the system is determined by the density of the universe
  at the time at which the system collapsed (or underwent its last major
  merger), so the older systems have the highest density and are therefore
  the hottest.
\item Supernova heating, though not the dominant source of energy in the
  halos, does have a second order effect on them. Perhaps in lower mass
  systems, the relatively higher contribution from SNIa causes gas to be
  lost from the potential, lowering the overall fraction.
\end{enumerate}

Of these suggested models, the last seems the least likely. Although it is
plausible that supernova heating could have observable effects on the
galaxy halos, particularly in the least massive members of our sample, it
seems unlikely that it could affect the gas fraction in the same way at all
radii. In the analogous situation of gas heating in galaxy groups, the
central gas fraction is lowered, but only by moving gas to higher radii,
raising the gas fraction there. There are however some obvious differences
which could make this comparison invalid, such as the distributed heating
and mass injection expected from the stars in a galaxy. It is also
important to remember that we cannot in most cases observe the galaxy halo
to high radii. It is possible that gas could be removed to a large fraction
of the virial radius, where its low density might make it undetectable, and
where it could be easily stripped by any interaction with a surrounding
IGM. However, the fact that we see no sign of any trend in \Bfit\ is a
strong argument against this hypothesis.

There are also problems with the idea that such a trend might be a natural
result of galaxy halos forming from gas of stellar origin. Previous work on
a larger sample including galaxies found in a variety of environments and
with a wide range of luminosities suggests that galaxy winds are the most
likely formation method for the halos of early--type galaxies of all sizes
\citep{OFP01age,OFP01cat}. However, these studies also confirmed that large
group or cluster dominant ellipticals, such as those in our sample, have
X--ray properties which differ significantly from those of less massive
galaxies. They could therefore be the most likely group of objects to form
their gas halos through some other mechanism. Our results suggest that the
steep trend in the \LT\ relation is caused by the drop in \Fgas\ and hence
\Lx. The \LxLbtwo\ relation also has a steep slope compared to that of
non--BGG ellipticals \citep{OFP01cat}, and it seems likely that this is
again a product of the trend in \Fgas\ with temperature. If so, we cannot
argue that the trend in gas fraction is a product of halo formation by
galaxy winds, as we would expect all galaxies to show a steep \LxLbtwo\ 
relation, as seen for this sample.

A change in gas fraction caused by the different formation histories
of the galaxies seems a more feasible model. Assuming our galaxies
have formed through a number of mergers, we can assume that there is a
scatter in the time since the last major merger. While minor mergers
can be treated as the accretion of extra matter into the galaxy, major
mergers will have more profound effects, including the ``resetting''
of the potential well through violent relaxation
\citep{Salvadoretal98} and the fuelling of bursts of star
formation. Luminosity weighted ages (based on comparing optical
absorption line indices to stellar population models) tend to be
dominated by any young component in the stellar population, so that if
the galaxy has formed part of its stellar population through a merger
and starburst, the age measured is likely to be an estimate of the
time since that merger. In a previous paper \citep{OFP01age} we found
a relation between the luminosity weighted age of a sample of 77
early--type galaxies in a range of environments and their normalised
X--ray luminosity, \LxLb. This indicates that the halo luminosity of
such galaxies increases as they age. In galaxy wind models, this can
be explained through the transition from rapid outflows to slower
winds or inflows, during which the halo density and the gas mass in
the potential well increases. 

The question then becomes whether we expect the more massive galaxies in
this sample to have had their last major merger earlier, and therefore to
have higher halo densities. Hierarchical models of galaxy formation predict
that more massive galaxies have longer formation timescales
\citep{Kauffmann96}, suggesting that this is not the case. However,
observational measurements of age for sample of galaxies in clusters and
the field suggest that more massive galaxies do indeed form earlier
\citep{Thomasetal02}. In our previous study of X--ray variations with
galaxy age we found no correlation between \LB\ and age \citep{OFP01age},
but in that case our sample included galaxies from a much wider range of
environments and with more varied luminosities than is the case in this
paper. In the current sample only 16 of our galaxies have measured ages in
the catalogue of \citet{Terlevichforbes00}, and for this subsample we find
only a weak (1.4 $\sigma$) trend for age to increase with \LB. We are
therefore unable to say whether merger history is the deciding factor in
determining gas fraction, though it does seem possible.

Finally there is the possibility that the change in gas fraction we see is
related to the environment of the galaxies. The offsets we see in the
\LxLbtwo\ and \LbT\ relations suggest that the presence of a surrounding
IGM can increase the X--ray luminosity and temperature of the halo. \Lx\ 
appears to be more strongly affected, being offset by a factor of $\sim$9,
as compared to a factor of $\sim$2 in \Tx. A surrounding IGM might cause
these offsets by containing or stifling galaxy winds, or by through
accretion of the IGM on to the galaxy halo. If the most massive galaxies
were found in the most massive groups and clusters we might expect the
density of the surrounding IGM to correlate with galaxy mass. This would
mean that more massive systems were more strongly affected, and might
produce the trend in gas fraction with temperature. Of the possible
processes, containment or stifling would seem the most likely, as accretion
of IGM gas might be expected to affect gas fraction differently at
different radii. One strong argument for this explanation is the correlation
between group and galaxy X--ray luminosity observed by
\citet{Helsdonetal01}. If we assume that this is not caused by group
cooling flows then it does indeed suggest that galaxy mass is correlated
with group mass.

\subsection{Environmental Effects and Formation Epoch}
Although most of the relations which we have used to investigate our sample
of galaxies show no difference between galaxies in different environments,
there are three interesting exceptions. These are the \LbT, \LxLbtwo\ and
\BfitT\ relations. The \LbT\ and \LxLbtwo\ both show similar behaviour. In
each case, we find that the best fit relation has a similar slope to that
of groups, but is offset to higher \Lx\ or \Tx\ at a given \LB. However,
when we split the sample into galaxies which lie in X--ray bright or X--ray
faint groups, we find that those in X--ray faint groups do not appear to be
offset to the same degree. We have calculated the residuals about our best
fit lines for both relations, and the residuals we would find if the best
fit lines for galaxy groups were used. In both cases, the mean residual
from our best fit line is considerably higher (by 20-30\%) for the galaxies
in X--ray faint groups than for those in X--ray bright groups, showing them
to be poorly described by the relations. The points are almost all offset
toward the group relations, so these large residuals are a measure of that
offset. We can also calculate the mean residual from the group relations,
which will give us a measure of the offset of the X--ray faint and bright
subsets from those relations. The mean residual from the best fit line for
groups is smaller (by 40-60\%) for galaxies in X--ray faint groups. For the
\LxLbtwo\ relation, the mean residual of the galaxies in X--ray faint
groups from the \emph{group} relation is actually less than half the mean
residual from the \emph{galaxy} best fit line, suggesting that although
these galaxies are not well described by either relation, they are
closer to the group relation. This supports the idea that the
offset is driven by the presence of a surrounding IGM.  Alternative
subdivisions of the sample, comparing cooling flow and isothermal galaxies,
or group dominant galaxies to those in other environments do not show a
similar segregation, leading us to the conclusion that this difference is
not a product of group cooling flows.

There are a number of ways in which a surrounding IGM could increase the
luminosity and temperature of a galaxy halo. If we consider a galaxy in an
early wind phase, the rapidly outflowing gas might expand until stopped by
the pressure of the surrounding gas. The IGM could contain the galaxy wind,
raising its density and preventing the dispersion of high temperature gas
to large radii. We might therefore expect such an object, even if it has
now progressed to a later wind phase, to have a higher
temperature and luminosity. Similarly, if we consider a galaxy in which
there is a global inflow (within the galaxy halo rather than a group
cooling flow initially sourced in the surrounding IGM), we might expect gas
from the IGM to move into the galaxy potential. This gas would add to the
galaxy luminosity, particularly as the galaxy potential would concentrate
it. We might also expect it to have a higher temperature than gas produced
within the galaxy. In
both cases we might also see an increased amount of gas in the galaxy at all
radii. If the trend in gas mass fraction discussed in
Section~\ref{sec:disc_LT} is caused by an increase in halo density for high
temperature systems (rather than a density decrease in low \Tx\ galaxies),
these models might give us an explanation of the trend. More massive
galaxies would produce larger amounts of gas and would be more able to
accrete and concentrate a surrounding IGM, so we would expect the trend
with temperature.

The other sign of environmental influence on our galaxies is shown in
Figure~\ref{fig:BetaT_CF}. If we exclude NGC~1404 (where \Bfit\ is likely
to be biased by the effects of ram-pressure stripping), we find that
galaxies in the centres of groups, clusters and cluster sub-clumps show a
larger degree of scatter in \Bfit\ than those elsewhere. In this case the
trend seems to be related to the position of the galaxy at the centre of
the potential well rather than a surrounding or confining IGM. All the
non-BGG/BCG galaxies in our sample are found in groups or clusters, so
their halos are likely to interact with surrounding gas. A possibility is
that the halos of galaxies at the centres of larger structures are affected
by the group potential, rather than the IGM. A recent study of NGC~1399
\citep{Paolilloetal02} shows that in the central regions of that galaxy
(the dominant elliptical of the Fornax cluster), the mass profile is
dominated by the stellar mass of the galaxy. However, the cluster potential
does play some part at larger radii, and although we have removed the
surface brightness components associated with larger structures, we cannot
rule out some level of influence. Another possibility is that the merger
history of these central galaxies has an effect on their halo structure.
Dynamical friction is thought to cause galaxies in groups and clusters to
lose orbital energy and fall towards the centre of the potential. We might
therefore expect to find that group and cluster central galaxies undergo
many more mergers than galaxies elsewhere. The associated
disturbance, influx of gas, and star formation might produce effects which
can change the state of the galaxy halo, inducing the scatter we observe.

Although environmental influences seem likely, and provide an explanation
for the offsets seen in the \LbT\ and \LxLbtwo\ relations, this model
raises further questions, when compared to the other relations. There is a
discrepancy between the \sigT\ and \LbT\ relations. In the \sigT\ relation,
temperature appears to be a good measure of mass, with a behaviour similar
to that seen in galaxy clusters. In the \LbT\ relation we see an offset,
suggesting that while \Tx\ may be a good measure of mass, either galaxies
have higher temperatures than clusters at a given mass, or that they have
lower optical luminosities. If we accept the \sigT\ relation, we must
assume that the offset is in \LB, and at first glance this suggests that
galaxies have a higher mass-to-light ratio than clusters. This is rather
unexpected, as cosmological simulations suggest that mass-to-light ratio is
correlated with dark halo mass, with galaxy clusters having considerably
higher mass per unit \LB\ than individual galaxies \citep{Somervilleetal01}.
A similar discrepancy is seen between the \LT\ and \LxLbtwo\ relations, in
which the former has an offset to lower \Lx\ (or higher \Tx), while the
latter has an offset to higher \Lx\ (or lower \LB). Once again, if we
assume temperature behaves as it does in clusters, then the \LT\ relation
suggests that \Lx\ is reduced, relative to larger structures. The offset in
the \LxLbtwo\ relation would then again imply a large reduction in \LB,
compared to groups and clusters. It is not clear what is causing these
offsets, but we can speculate to some extent.

One possibility to be considered is the difference in formation epoch
between galaxies, groups and clusters. Although properties such as $\sigma$
and X--ray temperature can be considered as measures of system mass, they
are actually related to the depth of the potential well
(\Tx~$\propto$~M/R). We expect systems to have densities related to the
critical density of the universe at the time at which they virialise (or
undergo their last major merger), so systems of a given mass which form at
earlier epochs should have deeper potentials than their more modern
counterparts. This suggests that older systems will have higher velocity
dispersions, X--ray temperatures and luminosities than younger ones. A
comparison of systems of different ages should show these differences, and
relations involving these parameters would be affected them. Relations
involving optical luminosity might show them particularly well, if the
total stellar luminosity is related to system mass rather than the depth of
the potential well.

As we are considering similar systems at different redshifts, we take
the total mass, $M_{tot}$ to be constant, but the radius of the system is 
reduced at higher redshift, R $\propto$ (1+$z$)$^{-1}$. Therefore, density
scales as
\begin{equation}
\rho_{tot} \propto \frac{M_{tot}}{R^3} \propto (1+z)^3,
\label{eqn:rho}
\end{equation}
as expected. We also know that for bremsstrahlung radiation, the luminosity
can be written
\begin{equation}
L_X \propto \rho_{tot}^2R^3\sqrt{T_X}. 
\end{equation}
Since $\rho_{tot}$R$^3$ is the mass of gas in the system, we can rewrite
this as
\begin{equation}
L_X \propto (1+z)^3M_{tot}f_{gas}\sqrt{T_X},
\label{eqn:Lx}
\end{equation}
where \Fgas\ is the gas fraction.\\
\Tx\ is proportional to $M/R$, so
rearranging Equation~\ref{eqn:rho}, we can say that
\begin{equation}
T_X \propto M_{tot}^{\frac{2}{3}}(1+z),
\label{eqn:Tx}
\end{equation}
and substituting into Equation~\ref{eqn:Lx} will therefore give
\begin{equation}
L_X \propto (1+z)^{\frac{3}{2}}f_{gas}T_X^2.
\label{eqn:LxTx}
\end{equation}

We therefore expect galaxies formed at a redshift $z_{form}$ = 2 to have a
similar \LT\ relation to that of clusters, offset to higher luminosities at
a given \Tx\ by a factor of 3$^{\frac{3}{2}}$, and modified by the
behaviour of \Fgas. As we in fact see a steeper slope and an offset to
lower than expected luminosities, this suggests that \Fgas\ is both
dependent on the \Tx\ of the system (as we have demonstrated elsewhere,
\cf\ Figure~\ref{fig:SBprofs}) and 
in general lower than is the case in groups and clusters, countering the
expected offset to higher luminosities.

Other relations can also be considered in the same way. Velocity dispersion
scales with mass and radius in a similar way to temperature, $\sigma^2 \propto
M/R$. This means that we expect $\sigma^2$ and \Tx\ to scale with
redshift by the same factor, so regardless of formation epoch the \sigT\ 
relation should remain constant. This agrees well with our findings.
Optical luminosity is assumed to scale with total mass regardless of
formation epoch, in which case we can substitute it into
Equation~\ref{eqn:Tx}. Rearranging the equation shows that we would expect
a relation of the form \LB\ $\propto$ \Tx$^{1.5}$, like that of clusters,
but offset to lower luminosities by a factor of 3$^{\frac{3}{2}}$. This
prediction agrees less well with our results, as we find a slope somewhat
steeper than that of clusters, with an offset of a factor $\sim$3. We can
also substitute this predicted relation for \LbT\ into Equation~\ref{eqn:LxTx}
to find a prediction for the \LxLbtwo\ relation. This again predicts a
relation unlike that we observe, of the form \Lx\ $\propto$
\LB$^{\frac{4}{3}}$ with a large offset to high X--ray luminosities.
However, this relation contains an \Fgas\ term which is likely to steepen
the relation (as \Fgas\ is dependent on \Tx) and partially counter the
offset to high \Lx.
 
\subsection{3D models and mass estimates}
Gas fraction is clearly an important quantity for our galaxies. We can use
the three dimensional models of our galaxies (see Section~\ref{res:bfit})
to measure properties such as gas mass and total halo mass within a range
of radii, in much the same way as we calculate gas entropy at a tenth of
the virial radius. However it is important to emphasise that these models
are by necessity inaccurate. We expect errors in the virial radii of the
galaxies, as we do not know the true redshift of formation of each object.
We also expect errors associated with temperature structure in the galaxy
halos, as we assume all the halos to be isothermal. The values we find are
therefore of more interest as representing trends rather than as exact
measures. Table~\ref{tab:mass} lists mean and median values for our sample.
When calculating the means, we exclude four galaxies which we find to have
extreme values of gas fraction. In one case (NGC~4073) we find that the gas
mass exceeds the mass of the dark halo at all radii, suggesting that the
spatial fit is contaminated by group emission. The other three galaxies
(NGC~1332, NGC~1549, NGC~4261) all have gas fractions of $<$10$^{-5}$,
considerably lower than most galaxies in the sample. Although these are
extreme cases, they do demonstrate the hazard of accepting measurements
from these models at face value. The median values we calculate should be
unaffected by these extreme cases, but the very low median gas fraction and
very high median mass-to-light ratio within the virial radius make this
approach somewhat suspect. We have derived the mean gas fraction, stellar
mass fraction and total baryon fraction for the galaxies, assuming a
stellar mass-to-light ratio of 5 \Msol/\LB\ \citep{Pizzellaetal97}. We also
assume that the stellar component of the galaxy is contained within
$R_{virial}/3$.

\begin{table*}
\begin{center}
\begin{tabular}{lcccc}
\hline
 & \multicolumn{2}{c}{Mean} & \multicolumn{2}{c}{Median}\\
 & $R_{virial}$ & $R_{virial}/3$ & $R_{virial}$ & $R_{virial}/3$\\
\hline\\[-3mm]
 Total mass (\Msol) & 1.31$\times$10$^{13}$ & 4.25$\times$10$^{12}$ &
 1.21$\times$10$^{13}$ & 3.60$\times$10$^{12}$\\
 Gas mass (\Msol) & 2.39$\times$10$^{11}$ & 4.60$\times$10$^{10}$ &
 6.26$\times$10$^{10}$ & 1.89$\times$10$^{10}$\\
 Stellar mass (\Msol) & 4.75$\times$10$^{11}$ & 4.75$\times$10$^{11}$ &
 3.53$\times$10$^{11}$ & 3.53$\times$10$^{11}$\\
 Gas fraction & 0.019 & 0.012 & 0.006 & 0.004 \\
 Stellar fraction & 0.050 & 0.16 & 0.038 & 0.113\\
 Baryon fraction & 0.069 & 0.17 & 0.049 & 0.124 \\
 Star Formation Efficiency & 0.77 & 0.92 & 0.84 & 0.96\\
 Mass-to-Light ratio (\Msol/\LB) & 166.7 & 54.8 & 281.1 & 38.62 \\
\hline
\end{tabular}
\end{center}
\caption{\label{tab:mass}Mean and median masses and other derived quantities for our sample
  of galaxies.}
\end{table*}

The gas masses shown in Table~\ref{tab:mass} are quite large, as we would
expect given that most of our sample is made up of groups and cluster
dominant giant ellipticals. \citet{Bregman92} find X--ray gas masses for
their sample of elliptical galaxies in the range $\sim$10$^{8-11}$ \Msol.
Our values are comparable with the upper end of this range particularly if
we take in to account the effect of extrapolation out to the virial radius.
The mean mass-to-light ratio for the sample may be somewhat high.
\citet{Lloyd-DaviesPonman02} find the mean mass-to-light ratio of a sample
of 20 galaxy groups and clusters to be 120$\pm$20 \Msol/\LB, and other
comparable estimates include 88$\pm$33 \Msol/\LB\ \citep{EdgeStewart91} and
100 \Msol/\LB\ for the Perseus cluster \citep{Eylesetal91}. However, some
estimates suggest that both elliptical galaxies and galaxy clusters may
have mass-to-light ratios as high as 200 \Msol/\LB\ \citep{Bahcalletal95}.
Taking a conservative view, we assume that the total halo masses may be
overestimated to some degree, possibly owing to the influence of a
surrounding group or cluster halo on the galaxies. The mean gas, stellar
and baryon fractions will all be influenced by any overestimate of total
mass, so we expect our measured values to be lower than is really the case.
However, even given such a bias, it is clear that the mean gas fraction is
considerably lower than is usual in galaxy groups and clusters;
$\sim$1-2\%, as compared to $\sim$20\% in clusters
\citep{Lloyd-DaviesPonman02,MarkevitchVikhlinin97}. This is exactly the
kind of difference we expected to see, given the offsets to low luminosity
in the \LT\ and \LxLbtwo\ relations. 

The mean star formation efficiency
($M_*/[M_*+M_{gas}]$) is also very high compared to larger structures,
which typically have efficiencies of $\sim$0.2-0.3
\citep{Lloyd-DaviesPonman02,Cirimeleetal97,Arnaudetal92,Davidetal90}.  This
could suggest that galaxies convert much more of their gas into stars than
is the case in groups or clusters, an unsurprising result considering the
extent and temperature of a typical cluster halo. This result is also
consistent with theoretical modelling which suggests that galaxy sized
halos are likely to contain much more cool gas that larger systems
\citep{Baughetal99}. In ellipticals, cool gas is likely to have been formed
into stars during the merging process, leading to the high star formation
efficiencies we measure. However, high star formation efficiencies would
also be inferred if much of the gas has been removed from the systems via
processes such as galaxy winds. The mean baryon fraction for our sample is
quite low compared to measured values for galaxy clusters, which range from
0.11-0.3
\citep{Lloyd-DaviesPonman02,Zaroubietal01,Hradeckyetal00,Davidetal95}. The
low end of this scale is comparable with our value, given the large
uncertainties we expect in our result. However, for larger values of baryon
fraction in clusters we would have to conclude that up to $\sim$75\% of the
baryons originally in galaxies have been removed, probably being blown out
into the surrounding IGM. Once again, we emphasize that the values listed in
Table~\ref{tab:mass} cannot be taken as precise measurements. 

\section{Summary and Conclusions}
\label{sec:conc}
We have compiled a sample of 39 massive X--ray luminous early--type
galaxies for which there are long \rosat\ PSPC exposures
archived. Analysing these data we have carried out detailed spatial
and spectral fits, and extrapolated from these to approximate
three dimensional models of the galaxies. The properties measured from
these fits and models allow us to compare our sample to galaxy groups and
clusters. Galaxies may be comparable to these larger systems because at the 
simplest level they can be considered as dark matter halos containing hot
gas, much like the more massive structures in which many of them reside. If the
dark matter profile does not vary significantly with system mass, we might
expect the gas properties to be similar as well. To compare our galaxies
with larger galaxy systems we fit a number of relations which are commonly
used for groups and clusters, including the \LT, \sigT, \BfitT, \LbT\ and
\LxLbtwo\ relations. We are also able to subdivide our sample according to
environment (galaxies in the centres of groups/clusters, galaxies
surrounded by a dense IGM) and temperature structure (galaxies with
isothermal halos, central cooling or heating). As we have specifically
chosen our sample to include the most massive galaxies available, 
it is unsurprising that most of them are found to lie in the centres of
larger systems and that a majority show signs of central cooling.

We find a number of interesting correlations, and some
of the relations we fit show similarities to those found for groups
and clusters. These include:

\begin{enumerate} 
\renewcommand{\theenumi}{(\arabic{enumi})} 
\item The \sigT\ relation is identical (within errors) to that found
for galaxy clusters, suggesting that both gas temperature and central
velocity dispersion are good measures of the depth of the potential
well. In clusters this correlation arises because gas is heated
through shocks as it falls into the potential and this may also be the
case for our galaxies. However, it also seems possible that
temperature could be related to the depth of the potential if the gas
is mainly of stellar origin, at least in large galaxies such as those
in our sample.

\item The \LT\ relation has a very similar slope to that found for
galaxy groups. Although this could be caused if group scale cooling
flows are the main source of the halos of our sample of galaxies, the
galaxies in which we find evidence of cooling do not behave
differently from those which show no signs of cooling, in this or any
of our other relations. This suggests that cooling flows are not the
predominant source of emission in our sample of galaxies.

\item The \LxLbtwo\ and \LbT\ relations also have similar slopes to
those found for galaxy groups.

\item The X--ray surface brightness profiles of the galaxies in the sample
have a mean \Bfit\ parameter of $\sim$0.5, similar to the mean optical
surface brightness slope expected for elliptical galaxies. This
suggests that there is equipartition of energy between the gas and
stellar components of these systems, a conclusion we also draw from
the slope of the \sigT\ relation. However, there is a variation in the
normalisation of the profiles with temperature, indicative of a
variation in the overall gas fractions of the systems. This variation
is likely to be the cause of the steep slope of the \LT\ relation,
with lower temperature systems having lower gas fractions and
therefore lower X--ray luminosities.
\end{enumerate}

We also find a number of differences between the relations found for
galaxies and those known to exist in groups and clusters, such as:

\begin{enumerate} 
\renewcommand{\theenumi}{(\arabic{enumi})} 
\item The \LT, \LxLbtwo\ and \LbT\ relations, which have similar
slopes to the relations found for galaxy groups, are offset from those
relations. These offsets could be at least partially explained if the
halos of the galaxies were in fact group cooling flows, but as
mentioned above this appears to be unlikely.

\item There is no trend in mean gas entropy with temperature, which we
might expect to find if galaxy halos were formed in the same way as
those of galaxy groups and clusters. 

\item There is also no trend in surface brightness slope with
temperature. In groups this trend is caused by the heating and
movement of gas to higher radii; in galaxies our results suggest that
this does not occur. Instead we find a trend in overall gas fraction
with temperature. Could this be a sign that heating expels gas from
the system, or that the halo builds up to higher densities and
temperatures over time?

\item In the \LbT\ and \LxLbtwo\ relations we find some evidence that
galaxies in groups whose X--ray halos are poor do not follow the same
relations as those galaxies which are surrounded by a rich IGM. This
may be evidence of environmental influence on our galaxies, though it
does not appear that this effect is caused by cooling flows. 
\end{enumerate}

Taken as a whole, our results seem to suggest that although galaxy
X-ray halos obey relations similar to those found for groups and
clusters, they are not formed through the same process. It seems
likely that a significant portion of the gas in the halos of
early--type galaxies is produced by stellar mass loss within the
galaxy, rather than infall and shock heating of primordial
material. The similarities in the behaviour of the halos across the
range of system sizes would in this case be a product of the dark
matter potential, which is believed to follow a profile of similar form, 
regardless of system mass. Factors such as the formation and
merger history of the galaxy, gas cooling and the surrounding
environment are all likely to influence the development of the halo,
and further exploration of the properties of early--type galaxies using
higher quality X--ray data may allow some of the outstanding issues
raised by this study to be resolved. It is particularly important to
remember that a large proportion of the galaxies in this sample lie at
or near the centres of galaxy groups. The advent of \chandra\ and
\xmm\ makes an improved study in this area possible, once a sufficient
number of observations of X--ray bright early--type galaxies in a
range of environments become available.

\vspace{1cm}
\noindent{\textbf{Acknowledgements}}
The authors would like to thank A. Sanderson, S. Helsdon and D. Forbes for
numerous helpful discussions and the benefit of their advice regarding
software. We also thank an anonymous referee for their comments, which have
improved the paper. This research made use of the NASA/IPAC Extragalactic
Database (NED, which is operated by the Jet Propulsion Laboratory,
California Institute of Technology, under contract with the National
Aeronautics and Space Administration), the LEDA database
(http://leda.univ-lyon1.fr), and data obtained from LEDAS (the Leicester
Database and Archive Service at the Department of Physics and Astronomy,
Leicester University, UK). The authors made use of Starlink facilities at
the University of Birmingham. E.O'S.  acknowledges the receipt of a PPARC
studentship and a grant from the University of Birmingham Caroline Harold
Research Fund.

\bibliographystyle{mn2e}
\bibliography{../paper}

\label{lastpage}

\end{document}